\documentclass[12pt, amsfonts, epsfig, reqno, a4paper]{amsart}
\usepackage[hyperindex,breaklinks]{hyperref}
\usepackage{amsmath}
\usepackage{amsfonts}
\usepackage{mathrsfs}
\usepackage{graphicx}
\usepackage{color}
\usepackage{slashed}
\usepackage{braket}
\usepackage{extarrows}
\usepackage{amsmath, bm}
\usepackage{amssymb}
\usepackage[margin=1in]{geometry}
\usepackage{bbm}
\usepackage{hyperref}

\usepackage{tikz}
\usepackage{nicematrix}

\newcommand{\half}{\frac12}

\definecolor{yellow}{rgb}{1,.7,0}

\definecolor{pkured}{rgb}{0.55,0,0}

\newcommand{\be}{\begin{equation*}}
	\newcommand{\ee}{\end{equation*}}
\newcommand{\beq}{\begin{equation}}
	\newcommand{\eeq}{\end{equation}}

\numberwithin{equation}{section}

\newtheorem{cor}{Corollary}[section]

\newtheorem{lem}[cor]{Lemma}
\newtheorem{prop}[cor]{Proposition}
\newtheorem{thm}[cor]{Theorem}

\newtheorem{ex}[cor]{Example}
\theoremstyle{remark}
\newtheorem{rmk}[cor]{Remark}

\numberwithin{figure}{section}
\usepackage{ytableau}

\usepackage{tikz}
\newcounter{x}
\newcounter{y}
\newcounter{z}

\newcommand\xaxis{210}
\newcommand\yaxis{-30}
\newcommand\zaxis{90}

\newcommand\topside[3]{
	\fill[fill=white, draw=black,shift={(\xaxis:#1)},shift={(\yaxis:#2)},
	shift={(\zaxis:#3)}] (0,0) -- (30:1) -- (0,1) --(150:1)--(0,0);
}

\newcommand\leftside[3]{
	\fill[fill=white, draw=black,shift={(\xaxis:#1)},shift={(\yaxis:#2)},
	shift={(\zaxis:#3)}] (0,0) -- (0,-1) -- (210:1) --(150:1)--(0,0);
}

\newcommand\rightside[3]{
	\fill[fill=white, draw=black,shift={(\xaxis:#1)},shift={(\yaxis:#2)},
	shift={(\zaxis:#3)}] (0,0) -- (30:1) -- (-30:1) --(0,-1)--(0,0);
}

\newcommand\cube[3]{
	\topside{#1}{#2}{#3} \leftside{#1}{#2}{#3} \rightside{#1}{#2}{#3}
}

\newcommand\planepartition[1]{
	\setcounter{x}{-1}
	\foreach \a in {#1} {
		\addtocounter{x}{1}
		\setcounter{y}{-1}
		\foreach \b in \a {
			\addtocounter{y}{1}
			\setcounter{z}{-1}
			\foreach \c in {1,...,\b} {
				\addtocounter{z}{1}
				\cube{\value{x}}{\value{y}}{\value{z}}
			}
		}
	}
}

\author{Chenglang Yang}
\email{yangcl@amss.ac.cn}
\address{Chenglang Yang, Hua Loo-Keng Center for Mathematical Sciences,
	Academy of Mathematics and Systems Science,
	Chinese Academy of Sciences,
	Beijing, China}

\title[{A remark on certain restricted plane partitions and crystal melting model}]
{A remark on certain restricted plane partitions and crystal melting model}

\begin{document}
\maketitle
%%%%%%%%%%%%%%%%%%%%%%%%%%%%%%%%%%%%%%%%%%%%%%%%%%%%%%%%%%%%%%%%%%%%%%%%%%%%%%%%%%%

\begin{abstract}
	In this paper,
	we provide formulas to calculate the partition functions of two types of plane partitions using the crystal melting model introduced by Okounkov, Reshetikhin and Vafa.
	As applications,
	we obtain a product formula for the partition function of the plane partitions with a limit boundary.
	A corollary of this formula is the demonstration of the equivalence between
	this partition function and the open string amplitude of the double-$\mathbb{P}^1$ model.
	We also derive a product formula for the partition function of symmetric plane partitions with a limit boundary along the $z$-axis direction.
\end{abstract}

\setcounter{section}{0}
\setcounter{tocdepth}{2}

%\tableofcontents

\section{Introduction}

Plane partitions are planar analogs of the ordinary integer partitions,
so they are also called the 3-dimensional partitions.
Intuitively,
they can be visualized as a collection of cubes piled in a corner.
The study of plane partitions in mathematics was initiated by MacMahon around 1900.
He obtained the following partition function of plane partitions in a box (see, for example, \cite{Mc})
\begin{align}\label{eqn:macmahon}
	M(q):=
	\sum_{\pi\in\mathcal{P}(a,b,c)} q^{|\pi|}
	=\prod_{i=1}^a \prod_{j=1}^b \frac{1-q^{i+j+c-1}}{1-q^{i+j-1}},
\end{align}
where $\mathcal{P}(a,b,c)$ denotes the set of plane partitions in an $a\times b\times c$ box,
and $|\pi|$ represents the size of the plane partition $\pi$.
Subsequently,
various methods were applied to this problem,
and numerous special types of plane partitions were introduced
(see \cite{Mac, S99}, see also \cite{ORV, FW07, V07, V09}).

The melting crystal model was introduced by Okounkov--Reshetikhin--Vafa \cite{ORV} when they studied the connections between the topological vertex in local Calabi--Yau geometry \cite{AKMV,LLLZ} and plane partitions.
They showed that,
the partition function of plane partitions with certain limit boundary conditions is equivalent to the topological vertex.
Furthermore,
both the partition function and the topological vertex admit formulas involving vacuum expectation value.

In Okounkov--Reshetikhin--Vafa's paper,
they introduced the so-called perpendicular partition function.
This partition function serves as the generating function for plane partitions inside a box with certain perpendicular boundary conditions (see Section 3.4 in \cite{ORV}).
Using the transition matrix method and vertex operators,
they derived a vacuum expectation value formula for this perpendicular partition function.
Notably,
their derivation assumes an additional condition that the height of the box is infinite.
However, their formula does not inherently enumerate such plane partitions.
Their formula counts the diagonal plane partitions up to a global correction factor (equation (3.15) in \cite{ORV}).
This correction term does not reconcile the difference between the partition functions of these two types of plane partitions. 
Subsection \ref{sec:ex} in this paper provides a straightforward example to illustrate and distinguish between them.
In this paper,
also in terms of the method introduced by Okounkov--Reshetikhin--Vafa,
we study these two types of plane partitions 
and provide formulas for their partition functions.

\begin{thm}\label{thm:main}
	Denote by $Z^{L,N,M}_{\lambda,\mu,\nu}$
	and $\tilde{Z}^{L,N,M}_{\lambda,\mu,\nu}$
	the partition functions of perpendicular plane partitions and diagonal plane partitions respectively.
	We have
	\begin{align}\label{eqn:main Z}
		\begin{split}
		&Z^{L,N,M}_{\lambda,\mu,\nu}
		=\delta_{L>\mu_1}\delta_{M>\mu^t_1}
		\cdot \tilde{\delta}_{L,N,\mu,\lambda^t}
		\tilde{\delta}_{M,N,\mu^t,\nu}
		\cdot q^{L\lambda^t_1+N|\mu|+M\nu_1}
		\cdot q^{-\lambda^t_1/2-\nu_1/2}\\
		&\quad\quad\cdot\langle (\lambda^t_1)|
		\prod_{0\leq j<L-1}^{\longleftarrow} \Gamma^{N-1,(k'_j,\lambda^t_{L-j})}_{+,\{j,\mu\}}(q^{j+\half})
		\cdot \mathbbm{1}_{l(\cdot^t)\leq N-1}
		\cdot \prod_{0\leq i<M-1}^{\longrightarrow} \Gamma^{N-1,(k_i,\nu_{M-i})}_{-,\{i,\mu\}}(q^{i+\half})
		|(\nu_1)\rangle
		\end{split}
	\end{align}
	and
	\begin{align}\label{eqn:main tZ}
		\begin{split}
		\tilde{Z}^{L,N,M}_{\lambda,\mu,\nu}
		=&\delta_{L>\mu_1}\delta_{M>\mu^t_1}
		\cdot q^{L|\lambda|+N|\mu|+M|\nu|}
		\cdot q^{-|\lambda|/2-|\nu|/2
			-\binom{\lambda}{2}-\binom{\nu^t}{2}}\\
		&\cdot\langle \lambda^t|
		\prod_{0\leq j<L-1}^{\longleftarrow} \Gamma^{N-1}_{+,\{j,\mu\}}(q^{j+\half})
		\cdot \mathbbm{1}_{l(\cdot^t)\leq N-1}
		\cdot \prod_{0\leq i<M-1}^{\longrightarrow} \Gamma^{N-1}_{-,\{i,\mu\}}(q^{i+\half})
		|\nu\rangle.
		\end{split}
	\end{align}
	The operators $\Gamma^{N-1,(a,b)}_{\pm,\{j,\mu\}}(q^{j+1}), \Gamma^{N-1}_{\pm,\{j,\mu\}}(q^{j+1})$ and $\mathbbm{1}_{l(\cdot^t)\leq N-1}$ used in above equations consist of vertex operators and projection operators.
	See Section \ref{sec:main} for more details. 
\end{thm}

In general,
it is not easy to compute the above vacuum expectation values directly.
This is mainly due to the complexity involved in handling the projection operator in the Fock space.
A very non-trivial example of physically dealing with such a problem can be found in \cite{Oku05} under Chern--Simons theory.
However,
despite the difficulties,
the formulas \eqref{eqn:main Z} and \eqref{eqn:main tZ} remain effective in some interesting cases,
including the example presented in Subsection \ref{sec:ex},
as well as in the computation of the partition function of plane partitions with a limit boundary discussed in Section \ref{sec:case1}.

The equation (3.21) in \cite{ORV} showed the equivalence between the topological vertex and the perpendicular partition function when the size of the box is infinite ($L,N,M$ go to infinity).
Their result is based on the equation (3.17) which calculates the diagonal partition function with two finite walls.
However,
as we will show in subsection \ref{sec:ex},
these two types of partition functions are not equal when $L, M$ are finite.
In the following proposition, we show that by letting $L,N,M\rightarrow\infty$,
the partition functions of these two types of plane partitions are equal.
More details can be seen in Section \ref{sec:main}.

\begin{prop}\label{prop:main equivalence}
	When $L=N=M=\infty$,
	\begin{align}
		\tilde{Z}^{\infty,\infty,\infty}_{\lambda,\mu,\nu}
		=Z^{\infty,\infty,\infty}_{\lambda,\mu,\nu}
		\in\mathbb{Z}[\![q]\!][q^{-1}].
	\end{align}
\end{prop}

As an application of the formula \eqref{eqn:main tZ} mentioned above,
we study a kind of plane partition which has a limit boundary along the $z$-axis direction.
The result is a product formula for the partition function of these plane partitions.

\begin{thm}\label{thm:main ex}
	The partition function $\tilde{Z}^{N,\infty,L}_{\emptyset,\mu,\emptyset}$ of plane partitions, bounded by two walls in two directions and admitting a certain limit boundary described by $\mu$ in the third direction,
	has the following product formula
	\begin{align}\label{eqn:Z^(N,infty,L) main}
		\tilde{Z}^{N+1,\infty,L+1}_{\emptyset,\mu,\emptyset}
		=\delta_{N\geq\mu_1} \delta_{L\geq\mu^t_1}
		\cdot\prod_{1\leq n\leq N \atop 1\leq l\leq L}
		(1-q^{n+l-1})^{-1}
		\cdot
		\frac{\prod_{(i,j)\in\mu} (1-q^{N-c(i,j)})
			(1-q^{L+c(i,j)})}
		{\prod_{(i,j)\in\mu} (1-q^{h(i,j)})}.
	\end{align}
\end{thm}

The product-type formula \eqref{eqn:Z^(N,infty,L) main} bears a resemblance to those partition functions of other types of plane partitions,
such as the ordinary plane partitions in a box \cite{Mc, Mac, S99},
the symmetric plane partitions and the shifted plane partitions \cite{Mc99,A78,Mac,S99,FW07,V07,V09}.
In each of those cases,
the plane partition is either confined in a box or has no specific boundary conditions.
In contrast, our case considers the plane partitions with a
certain limit boundary along the $z$-axis direction.
Thus, it is somehow interesting that a product formula still exists for the partition function of such plane partitions.
Notably, for those examples \cite{Mc,Mac,S99,A78,V09},
pure combinatorial proofs exist for their product formulas.
Providing a pure combinatorial proof for our formula \eqref{eqn:Z^(N,infty,L) main}
becomes an interesting and potentially difficult question.
Essentially,
the plane partitions with a limit boundary are equivalent to the skew plane partitions in \cite{OR07}.
Thus, this model is a special case of the Schur process introduced in \cite{OR03}.
The corresponding partition function formula was obtained in the second equation of Theorem 2 in \cite{OR07}.
Their formula is a product of terms labeled by elements in two sets.
Notably,
the main terms in our formula \eqref{eqn:Z^(N,infty,L) main} are a special case of MacMahon formula \eqref{eqn:macmahon} and a product of terms labeled by boxes in the Young diagram corresponding to $\mu$.
It is not apparent that these two kinds of formulas are the same even they should be equal to each other.
As we will see below,
the form in our formula can be used to study its relation to the double-$\mathbb{P}^1$ model and give a new proof of the full MacMahon formula.

The motivation behind deriving formula \eqref{eqn:Z^(N,infty,L) main} in this specific form is its connection to the open string amplitude of the double-$\mathbb{P}^1$ model.
The general philosophy (see \cite{Oku05,S06}) here is that
introducing a wall in the crystal melting model is equivalent to gluing a new topological vertex.
The closed string amplitudes of the resolved conifold, double-$\mathbb{P}^1$ and the closed topological vertex were studied in \cite{Oku05,S06}.
By comparing the formula \eqref{eqn:Z^(N,infty,L) main} with the open string amplitude of double-$\mathbb{P}^1$ with one nontrivial representation (see Subsection \ref{sec:dP} for a review),
we essentially provide an example of this philosophy within the context of open string amplitude.
\begin{cor}\label{cor:dP1}
	Denote by $Z^{double-\mathbb{P}^1}_{\mu;t_1,t_2}$ the open string amplitude of the double-$\mathbb{P}^1$ model with one nontrivial representation labeled by $\mu$.
	Ignoring the term $\delta_{N\geq\mu_1} \delta_{L\geq\mu^t_1}$ in formula \eqref{eqn:Z^(N,infty,L) main} for $\tilde{Z}^{N,\infty,L}_{\emptyset,\mu,\emptyset}$,
	we have
	\begin{align}
		\tilde{Z}^{N,\infty,L}_{\emptyset,\mu,\emptyset}
		=q^{-\|\mu\|^2/2}
		\cdot M(q)
		\cdot Z^{double-\mathbb{P}^1}_{\mu;t_1,t_2},
	\end{align}
	where $\|\mu\|^2=\sum_{i=1}^{l(\mu)}\mu_i^2$,
	$M(q)$ is the MacMahon formula \eqref{eqn:macmahon} and the K{\"a}hler parameters $t_1, t_2$ of double-$\mathbb{P}^1$ are determined by
	$e^{-t_1}=q^L, e^{-t_2}=q^N$.
\end{cor}

By employing the formula \eqref{eqn:Z^(N,infty,L) main} for the partition function of plane partitions with a limit boundary,
we also provide a proof of the full MacMahon formula.
The similar method has been previously applied in \cite{OR03,ORV} (see also \cite{V09}).
However,
it is worth noting that their method works only for the plane partitions without height restriction.
This is because the dependence of the formula on the vacuum expectation value on the $z$-axis direction is somewhat more complicated than the other two directions.
Our method works due to the closed formula presented in Theorem \ref{thm:main ex} and the rotation symmetry of plane partitions with perpendicular-type boundaries along three directions.
It will be a challenge to extend this method to partition functions considered in \cite{FW07,V07,V09} since there is no rotation symmetry in their cases.

We are also interested in the symmetric plane partitions.
MacMahon conjectured the first formula for the partition function of the symmetric plane partitions in 1899 \cite{Mc99},
and later it was proved in \cite{A78, Mac} (see also \cite{S99}).
In this paper,
we consider the extension to symmetric plane partitions that possess a limit boundary.
We derive a product formula for their partition function as follows
\begin{thm}\label{thm:SZ main}
	The partition function $SZ(N+1,\mu)$ of symmetric plane partitions,
	bounded by two walls and possessing a limit boundary described by $\mu$ along the $z$-axis direction,
	has the following product formula
	\begin{align}\label{eqn:SZ formula main}
		\begin{aligned}
			SZ(N+1,\mu)=&
			\delta_{\mu_1\leq N}
			\cdot\prod_{i=0}^{N-1}\frac{1}{(1-q^{2i+1})\prod_{j=0}^{i-1}(1-q^{2(i+j+1)})}\\
			&\quad\quad\quad\quad\cdot \frac{\prod_{(i,j)\in\mu}(1-q^{2N+2c(i,j)})}
			{\prod_{(i,i)\in\mu}(1-q^{h(i,i)})
				\prod_{(i,j)\in\mu\atop i<j}(1-q^{2h(i,j)})},
		\end{aligned}
	\end{align}
	where $\mu$ is a symmetric partition,
	$\delta_{\mu_1\leq N}=1$ if $\mu_1\leq N$ and otherwise it is $0$.
\end{thm}

The symmetric plane partitions are related to the free boundary Schur process (see \cite{BR05,BBNV}).
For example, in Section 5 of \cite{BBNV},
certain measure on the set of symmetric plane partitions is regarded as a special case of the free boundary Schur process and the Pfaffian formula for this process is applied to study the large random symmetric plane partitions.
The above formula \eqref{eqn:SZ formula main} for the partition function $SZ(N+1,\mu)$ may also be derived from Proposition 3.2 in \cite{BR05} after taking a special specialization in their formula.
Our formula mainly consists of two kinds of terms,
first of which is the MacMahon's partition function of symmetric plane partitions,
and second of which is a product of terms labeled by boxes in the Young diagram corresponding to $\mu$.

The rest of this paper is organized as follows.
In Section \ref{sec:pre},
we provide a review of the definitions and basic properties of plane partitions, Schur functions, vertex operators
and the double-$\mathbb{P}^1$ model.
At the end of this section,
we prove the Corollary \ref{cor:dP1}.
In Section \ref{sec:main},
by using the method of crystal melting model,
we prove Theorem \ref{thm:main}.
As applications,
we prove a product formula for the partition function of plane partitions possessing a limit boundary in Section \ref{sec:case1}.
Consequently,
we offer a new proof of the full MacMahon formula.
The similar methods are employed in Section \ref{sec:case2} to study the symmetric plane partitions possessing a limit boundary.

\section{Preliminaries}
\label{sec:pre}
In this section,
we review the plane partitions, Schur functions, vertex operators
and double-$\mathbb{P}^1$ model.
Most of them are necessary materials in the method of crystal melting model introduced by Okounkov--Reshetikhin--Vafa \cite{ORV}.

\subsection{Plane partitions}
In this subsection,
we review the notations for partition and plane partition for completeness.
For a reader who is not very familiar with these notations,
we recommend the Chapter I in \cite{Mac}.

\begin{figure}[htbp]
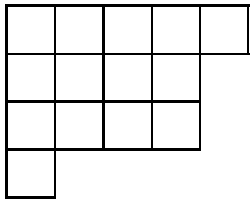

	\ydiagram{5,4,4,1}
	\caption{The Young diagram corresponding to $(5,4,4,1)$.}
	\label{eqn:Yd 5441}
\end{figure}

An ordinary partition of a nonnegative integer $n$ is a sequence of nonnegative weakly decreasing integers
\begin{align*}
	\mu=(\mu_1,\mu_2,...)
\end{align*}
satisfying the size of the partition $|\mu|:=\sum_{i=1}^\infty \mu_i=n$.
In general,
we can omit the zeros in a partition.
That is to say, a partition can be written as $\mu=(\mu_1,...,\mu_l)$ if $\mu_l\neq0$ and $\mu_{l+1}=0$.
The integer $l$ is called the length of the partition $\mu$.
Each partition has a Young diagram representation.
For example,
Figure \ref{eqn:Yd 5441} is the Young diagram corresponding to the partition $(5,4,4,1)$.
We will not distinguish $\mu$ and its corresponding Young diagram.
The partition $\mu^t$ is the conjugation of $\mu$ such that
$\mu^t_i=\#\{j|\mu_j\geq i\}$.
For example,
the conjugation of $(5,4,4,1)$ is $(4,3,3,3,1)$.
Intuitively, the Young diagram of $\mu^t$ is the transpose of $\mu$ along the main diagonal.

A partition could also be represented by its Frobenius notation,
\begin{align*}
	\mu=(m_1,...,m_{r(\mu)}|n_1,...,n_{r(\mu)}),
\end{align*}
where $m_i, n_j$ are integers satisfying $m_1>m_2>\cdots>m_{r(\mu)}\geq0$ and $n_1>n_2>\cdots>n_{r(\mu)}\geq0$.
To be precise,
$\mu$ and $(m_i,n_j)$ are determined by each other in terms of
\begin{align*}
	m_i=\mu_i-i, 
	\quad n_j=\mu^t_j-j,
	\quad 1\leq i,j \leq r(\mu),
\end{align*}
where $r(\mu)$ is the largest integer satisfying $(r(\mu),r(\mu))\in\mu$ when regarding $\mu$ as a Young diagram.
In general,
$r(\mu)$ is called the Frobenius length of $\mu$,
and it is intuitively the length of the diagonal of Young diagram $\mu$.
For example,
the Frobenius notation of partition in Figure \ref{eqn:Yd 5441} is $(4,2,1|3,1,0)$.
The content of $\mu$ at $(i,j)\in\mu$ is defined by $c(i,j)=j-i$
and the hook-length at $(i,j)$ is $h(i,j)=\mu_i+\mu^t_j-i-j+1$.

It is beneficial to introduce the notion of interlacing Young diagrams when studying the plane partition.
For two partitions $\mu$ and $\lambda$,
we say $\mu$ interlaces with $\lambda$ and write $\mu\succ\lambda$ if $\mu_j\geq\lambda_j\geq\mu_{j+1}$ for all $j\geq1$. 

A plane partition is a planar analog of the ordinary partition.
By definition,
a plane partition is a 2-dimensional sequence of nonnegative integers
\begin{align*}
	\pi=(\pi_{i,j}),
	\quad i,j=1,2,...
\end{align*}
satisfying the following weakly decreasing conditions
\begin{align}\label{eqn:def decreasing}
	\pi_{i+1,j}\leq\pi_{i,j}
	\quad \text{and} \quad
	\pi_{i,j+1}\leq\pi_{i,j}.
\end{align}
We call $\pi$ finite if the size of it $|\pi|:=\sum_{i,j}\pi_{i,j}$ is a finite integer.
Denote by $\mathcal{P}$ the set of all finite plane partitions.

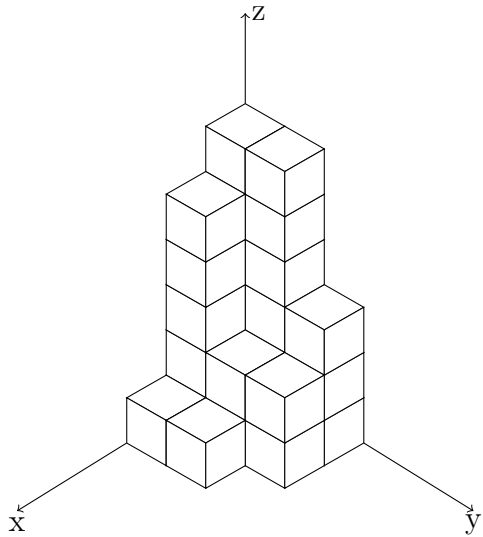
\begin{figure}[htbp]
	\begin{tikzpicture}[scale=0.5]
		\planepartition{{6,6,3},{5,2,2},{1,1}}
		\draw [->] (-2.5-0.1,-1.5)--(-5,-3);
		\node at(-5,-3.3) {x};
		\draw [->] (2.5+0.1,-1.5)--(5,-3);
		\node at(5,-3.3) {y};
		\draw [->] (0,6)--(0,8);
		\node at(0.3,8) {z};
	\end{tikzpicture}
	\caption{The 3D diagram corresponding to the plane partition $\pi$ in equation \eqref{eqn:pi ex}.}
	\label{figure:ex1}
\end{figure}

Intuitively,
a plane partition $\pi$ always corresponds to a 3D diagram such that the 3D diagram has $\pi_{i,j}$ cubes at the position $(x,y)\in[i-1,i]\times[j-1,j]$.
For example,
the corresponding 3D diagram of the following plane partition
\begin{align}\label{eqn:pi ex}
	\pi=\left(
	\begin{matrix}
		6 & 6 & 3 & 0 & \cdots\\
		5 & 2 & 2 & 0 & \cdots\\
		1 & 1 & 0 & 0 & \cdots\\
		0 & 0 & 0 & 0 & \cdots\\
		\vdots & \vdots & \vdots & \vdots & \ddots\\
	\end{matrix}\right),
\end{align}
where $\cdots$ are all zeros,
is given by Figure \ref{figure:ex1}.

For each plane partition $\pi$,
one can associate it with a sequence of interlacing Young diagrams defined by
\begin{align*}
	\mu^{k}:=\begin{cases}
		(\pi_{-k+i,i})_{i=1}^{\infty}  & \text{\ if\ } k<0 \\
		(\pi_{i,k+i})_{i=1}^{\infty}  & \text{\ if\ } k\geq0,
	\end{cases}
\end{align*}
where the interlacing Young diagrams mean that they satisfy
\begin{align*}
	\cdots\prec\mu^{-k}
	\prec\mu^{-k+1}
	\prec\cdots
	\prec\mu^{0}
	\succ \cdots
	\succ\mu^{k-1}
	\succ\mu^{k}\succ\cdots.
\end{align*}
Intuitively,
the $\mu^k$ is just the $k$-th diagonal slice of the plane partition.
More precisely,
the parts of $\mu^k$ correspond to the height of the columns of cubes in $k$-th slice of the plane partitions.
See Subsection 3.1 in \cite{ORV} for more details.
For example,
the corresponding interlacing Young diagrams of the plane partition in equation \eqref{eqn:pi ex} is
\begin{align*}
	\cdots, \mu^{-3}=\emptyset,
	\mu^{-2}=(1),
	\mu^{-1}=(5,1),
	\mu^{0}=(6,2),
	\mu^{1}=(6,2),
	\mu^{2}=(3),
	\mu^{3}=\emptyset,\cdots.
\end{align*}
Conversely,
a sequence of interlacing Young diagrams can also produce a plane partition,
and the interlacing condition between these Young diagrams corresponds to the weakly decreasing conditions \eqref{eqn:def decreasing} for the plane partition.
Thus, we can write
\begin{align*}
	\pi=(\mu^k)_{k=-\infty}^\infty.
\end{align*}
Obviously,
the finiteness of the plane partition $\pi$ corresponds to the condition $\sum_{k}|\mu^k|<\infty$.

\begin{rmk}\label{rmk:pi diagonal}
In general,
we will also consider the plane partitions in a restricted region.
For example,
in Section \ref{sec:main},
plane partition restricted in the region
\begin{align*}
	D(L,N,M)=\{(x,y,z)\in\mathbb{R}^3_{\geq0}|x-y\leq L, y-x\leq M, z\leq N\}
\end{align*}
will be considered,
where $L,N,M$ are some positive integers.
In this case,
the weakly decreasing condition in equation \eqref{eqn:def decreasing} will be relaxed if $(i,j)$ belongs to this region but $(i+1,j)$ or $(i,j+1)$ does not.
For example,
when $L=M=3, N=6$,
the following
\begin{align}\label{eqn:pi diagonal ex}
	\pi=
	\begin{pNiceMatrix}
		\Block[borders={right}]{1-3}{}6 & 6 & 4 & 0 & \cdots\\
		\cline{4}
		\Block[borders={right}]{1-4}{} 5 & 2 & 2 & 2 & \cdots\\
		\cline{5}
		2 & 1 & 0 & 0 & \cdots\\
		\cline{1}
		\Block[borders={right}]{1-1}{}0 & 1 & 0 & 0 & \cdots\\
		\cline{2}
		\Block[borders={right}]{1-2}{}\vdots & \vdots & \vdots & \vdots & \ddots\\
	\end{pNiceMatrix}
\end{align}
is still considered as a reasonable plane partition in the region $D(3,6,3)$.
\end{rmk}

\subsection{Schur functions}
\label{sec:pre schur}
We review the definitions and basic properties of Schur functions and skew Schur functions in this subsection.

The Schur function $s_\lambda$,
labeled by a partition $\lambda=(\lambda_1,...,\lambda_l)$ is a symmetric functions with respective to variables $x_i, 1\leq i<\infty$.
It is defined by
\begin{align*}
	s_\lambda
	=s_\lambda(x_1,x_2,...)
	=\lim_{n\rightarrow\infty}
	\frac{\det(x_i^{\lambda_j+n-j})_{1\leq i,j\leq n}}
	{\det(x_i^{n-j})_{1\leq i,j\leq n}}.
\end{align*}
By using the power sum coordinates
\begin{align*}
	p_k=p_k(x_1,x_2,...)=\sum_{i=1}^\infty x_i^k,
	\quad 1\leq k<\infty,
\end{align*}
the Schur function $s_\lambda$ is a homogeneous polynomial of degree $|\lambda|=\sum_{i=1}^{l(\lambda)}\lambda_i$
in the ring $\mathbb{C}[p_1,p_2,...]$
when assigning $\deg p_k=k$.

In this paper,
when considering the evaluation of Schur functions,
we always interpret them as functions of variables $x_i$ unless otherwise specified.
That is to say, for a sequence of numbers $(a_1,a_2,...)$,
the notation
\[s_\lambda(a_1,a_2,...)
=s_\lambda|_{x_i\rightarrow a_i}\]
denotes the evaluation of $s_\lambda$ at $x_i=a_i$.
For example,
denote by $\rho=(-1/2,-3/2,...)$
and $q^\rho=(q^{-1/2},q^{-3/2},...)$,
then
\begin{align*}
	s_\lambda(q^\rho)
	=s_\lambda(q^{-1/2},q^{-3/2},...)
	=s_\lambda|_{x_i\rightarrow q^{-i+1/2}}.
\end{align*}
The general evaluations of Schur function may make no sense since there are infinite variables.
However, the special evaluations in this paper are always meaningful in a suitable ring of formal power series.

Some special evaluations of Schur functions are well-studied and helpful in studying many problems (see the examples in I.3 in \cite{Mac}).
For a sequence $(a_1,a_2,...)$,
Denote by $(a_1,a_2,...)|_N=(a_1,...,a_N,0,0,...)$
the truncation of the original sequence.
Then by I.3.Example 1 in \cite{Mac},
\begin{align*}
	s_\lambda(q^{-\rho-1/2}|_N)
	=
	\delta_{l(\mu)\leq N}\cdot
	s_{\lambda}(1,q,q^2,...,q^{N-1},0,...)
	=\delta_{l(\mu)\leq N}\cdot q^{n(\lambda)}\prod_{(i,j)\in\lambda} \frac{1-q^{N+c(i,j)}}{1-q^{h(i,j)}},
\end{align*}
where $\delta_{l(\mu)\leq N}=1$ if $l(\mu)\leq n$, otherwise it is $0$,
and $n(\lambda)=\sum_i (i-1)\lambda_i$.
Just by the homogeneous condition,
\begin{align}\label{eqn:schur q^rho|N}
	s_{\lambda}(q^{-\rho}|_N)
	=\delta_{l(\mu)\leq N}\cdot s_{\lambda}(q^{1/2},q^{3/2},...,q^{N-1/2})
	=\delta_{l(\mu)\leq N}\cdot q^{n(\lambda)+|\lambda|/2}\prod_{(i,j)\in\lambda} \frac{1-q^{N+c(i,j)}}{1-q^{h(i,j)}},
\end{align}
and by letting $N\rightarrow\infty$,
\begin{align}\label{eqn:schur q^rho}
	s_{\lambda}(q^{-\rho})
	=s_{\lambda}(q^{1/2},q^{3/2},...)
	=q^{n(\lambda)+|\lambda|/2}\prod_{(i,j)\in\lambda} \frac{1}{1-q^{h(i,j)}}.
\end{align}
Note that,
denote by $\|\lambda\|^2=\sum_{i=1}^{l(\lambda)}\lambda_i^2$,
then
\begin{align*}
	n(\lambda)
	=\sum_{i=1}^{l(\lambda)}\sum_{j=1}^{\lambda_i} (i-1)
	=\sum_{j=1}^{l(\lambda^t)}\sum_{i=1}^{\lambda^t_j} (i-1)
	=\|\lambda^t\|^2/2-|\lambda|/2.
\end{align*}

There is a standard inner product on the ring $\mathbb{C}[p_1,p_2,...]$ such that
\begin{align*}
	( s_\lambda , s_\mu ) = \delta_{\lambda,\mu}
\end{align*}
for all partitions $\lambda,\mu$,
where $\delta_{\lambda,\mu}=1$ if $\lambda=\mu$, otherwise it is zero.
The skew Schur function is defined by
\begin{align*}
	s_{\lambda/\mu}=\sum_{\nu} c^{\lambda}_{\mu \nu} s_{\nu},
\end{align*}
where $\{c^{\lambda}_{\mu \nu}\}$ are the Littlewood--Richardson coefficients defined by $s_{\mu} s_{\nu}=\sum_{\lambda} c^{\lambda}_{\mu \nu} s_{\lambda}$.
Equivalently, the skew Schur functions are determined by the following equations
\begin{align*}
	(s_{\lambda/\mu},s_\nu)=(s_\lambda,s_\mu s_\nu)
\end{align*}
for all partitions $\nu$.
As a result,
$s_{\lambda/\mu}\in\mathbb{C}[p_1,p_2,...]$ is a homogeneous polynomial of degree $|\lambda|-|\mu|$.
A special evaluation of the skew Schur function is
\begin{align}\label{eqn:skew schur evaluate z}
	s_{\lambda/\mu}(1,0,0,...)=\begin{cases}
		0, & \lambda \nsucc \mu,\\
		1, & \lambda \succ \mu.
	\end{cases}
\end{align}

It will be much more convenient to use the following notation,
which comes from the boson-fermionic correspondence.
Since the fermions are not used in this paper,
we do not introduce the boson-fermionic correspondence
and just use its notation (see Appendix A in \cite{O01}).
For each partition $\mu$,
denote by $|\mu\rangle$ the associated Schur function $s_\mu$ in the ring $\mathbb{C}[p_1,p_2,...]$,
and the $\langle\mu|$ its dual.
Then,
for any differential operator $A$ over the ring $\mathbb{C}[p_1,p_2,...]$,
the action of $\langle\mu|$ on $A|\lambda\rangle$ can be written as the following form of the vacuum expectation value
\begin{align*}
	\langle\mu|A|\lambda\rangle
	=(|\mu\rangle, A|\lambda\rangle)
	=(A^*|\mu\rangle,|\lambda\rangle),
\end{align*}
where $\cdot$ means the action
and $A^*$ is the dual operator of $A$.
That is to say,
$\langle\mu|A$ could also be regarded as the dual of $A^* |\mu\rangle$.

When $\mu$ is the empty partition,
denote by the corresponding vector $|0\rangle=1\in\mathbb{C}[p_1,p_2,...]$ the constant function.
It is called the vacuum vector.
Its dual $\langle0|$ is called the dual vacuum vector.

\subsection{Vertex operators}
In this subsection,
we review the definition of vertex operators.
They provide an important tool to study Schur functions and the applications of Schur functions to many aspects of mathematics.
We mainly follow the notations in \cite{O01}.

Denote the Heisenberg operators by
\begin{align*}
	\alpha_n:=\begin{cases}
		n\frac{\partial}{\partial p_n}, &n>0\\
		p_{-n} \cdot, & n<0,
	\end{cases}
\end{align*}
where $p_{-n} \cdot$ is the operator multiplying by $p_{-n}$,
then the vertex operators are defined by
\begin{align*}
	\Gamma_\pm(z)=\exp\Big(\sum_{n=1}^\infty \frac{z^n \alpha_{\pm n}}{n}\Big).
\end{align*}
By the commutation relation $[\alpha_m,\alpha_n]=m\delta_{m+n,0}$,
one can prove that
\begin{align}\label{eqn:gamma+- comm relation}
	\Gamma_+(z)\Gamma_-(w)
	=\frac{1}{1-zw} \Gamma_-(w)\Gamma_+(z).
\end{align}
The above formula will be very useful when computing vacuum expectation value
and obtaining product-type formula.

Another useful differential operator over the ring $\mathbb{C}[p_1,p_2,...]$ is
\begin{align*}
	L_0:=\sum_{k=1}^\infty kp_k\frac{\partial}{\partial p_k}.
\end{align*}
It is just the homogeneous operator when assigning $\deg p_k=k$.
The commutation relation between $L_0$ and $\Gamma_\pm(z)$ is
\begin{align}\label{eqn:L0 gamma+_}
	q^{L_0} \ \Gamma_-(z)
	=\Gamma_-(qz) q^{L_0}
	\quad\text{and}\quad
	q^{L_0} \ \Gamma_+(z)
	=\Gamma_+(q^{-1}z) q^{L_0}.
\end{align}

The action of the vertex operators on Schur functions is given by the following,
\begin{align}\label{eqn:gamma- right}
	\Gamma_-(z)|\mu\rangle
	=\sum_{\lambda} s_{\lambda/\mu}(z,0,0,...) |\lambda\rangle
\end{align}
and its dual gives
\begin{align}\label{eqn:gamma+ left}
	\langle\mu|\Gamma_+(z)
	=\sum_{\lambda} s_{\lambda/\mu}(z,0,0,...) \langle\lambda|.
\end{align}
The other two cases are given by
\begin{align}\label{eqn:gamma+- action}
	\Gamma_+(z)|\mu\rangle
	=\sum_{\lambda} s_{\mu/\lambda}(z,0,0,...) |\lambda\rangle
	\quad\text{and}\quad
	\langle\mu|\Gamma_-(z)
	=\sum_{\lambda} s_{\mu/\lambda}(z,0,0,...) \langle\lambda|,
\end{align}
which can be simply proved by using the standard inner product.
The special cases of the above equations are
\begin{align}\label{eqn:gamma preserve vacuum}
	\Gamma_+(z)|0\rangle=|0\rangle
	\quad\text{and}\quad
	\langle0|\Gamma_-(z)=\langle0|.
\end{align}
The above two formulas are also evident from the definitions of the vertex operators and the vacuum vector $|0\rangle$.

A special case of the above equations \eqref{eqn:gamma- right} and \eqref{eqn:gamma+ left} are given by $z=1$.
In this case,
by the value of this special evaluation of skew Schur functions in equation \eqref{eqn:skew schur evaluate z},
we have
\begin{align}\label{eqn:gamma+- succ}
	\Gamma_-(1)|\mu\rangle
	=\sum_{\lambda\succ\mu} |\lambda\rangle
	\quad\text{and}\quad
	\langle\mu|\Gamma_+(1)
	=\sum_{\lambda\succ\mu} \langle\lambda|.
\end{align}
The above equations will be very useful for generating interlacing Young diagrams in studying plane partitions.

In general,
about the action of infinitely many vertex operators,
we also have (see, for example, equation (A.15) in \cite{O01})
\begin{align}\label{eqn:inf gamma+ action}
	\langle0|\prod_{i=1}^\infty \Gamma_+(z_i) |\mu\rangle
	=s_{\mu}(z_1,z_2,...).
\end{align}

\subsection{The open string amplitude of double-$\mathbb{P}^1$ model}
\label{sec:dP}
In this subsection,
we review the basic notation of the open string amplitude of double-$\mathbb{P}^1$ model following \cite{AKMV} and \cite{IK06}.
For the algebraic geometric side, we recommend \cite{LLLZ, MOOP}.
As a result,
we give a proof of the Corollary \ref{cor:dP1}.

\begin{figure}[htbp]
	\begin{tikzpicture}[scale=0.6]
		\draw  (0,0)--(3,0);
		\draw  (5,2)--(3,0);
		\draw  (3,-2)--(3,0);
		\draw  (0,3)--(0,0);
		\draw  (-3,-3)--(0,0);
		\draw  (-3,-3)--(-5,-3);
		\draw  (-3,-3)--(-3,-5);
		\node at(0.3,3) {$\mu$};
		\node at(1.35,0.4) {$Q_1$};
		\node at(-1.3,-2) {$Q_2$};
	\end{tikzpicture}
	\caption{The toric diagram of double-$\mathbb{P}^1$ with K{\"a}hler parameters $e^{t_{i}}=Q_{i}, i=1,2$.}
	\label{eqn:toric of dP}
\end{figure}
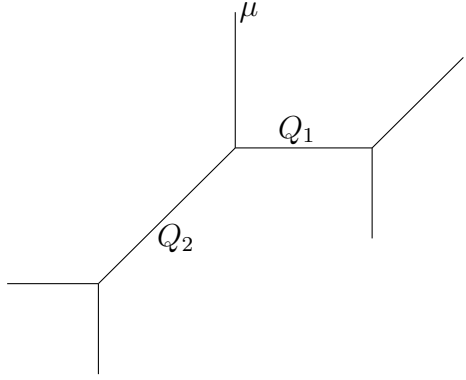

From the viewpoint of string theory,
the A-model topological string amplitudes of any Calabi--Yau threefolds are generating functions of the Gromov--Witten invariants of this manifolds \cite{LLLZ},
which should count maps from bordered Riemann surfaces to corresponding Calabi--Yau threefolds.
For the smooth toric Calabi--Yau threefolds,
a very explicit and effective method proposed by Aganagic, Klemm, Mari\~{n}o and Vafa is the topological vertex \cite{AKMV}
(see \cite{LLLZ} for a mathematical theory for the topological vertex).
Their method first gives an explicit formula \eqref{eqn:def W} for the topological vertex,
and then describes the gluing rules,
which express open Gromov--Witten invariants of general smooth toric Calabi--Yau threefolds in terms of the topological vertex.
To be precise,
the topological vertex is the generating function of open Gromov--Witten invariants of $\mathbb{C}^3$.
Let $\mu^1, \mu^2, \mu^3$ be three partitions.
They label the winding numbers of maps to $\mathbb{C}^3$ around three Lagrangian boundaries in $\mathbb{C}^3$.
Then a certain generating function of the open Gromov--Witten invariants of $\mathbb{C}^3$ labeled by $(\mu^1, \mu^2, \mu^3)$ is given by the following formula (see \cite{AKMV,LLLZ}, see also Proposition 4.4 in \cite{Z03} for the following explicit form)
\begin{align}\label{eqn:def W}
	\mathcal{W}_{\mu^1,\mu^2,\mu^3}(q)
	= (-1)^{|\mu^2|} q^{\kappa_{\mu^3}/2} s_{(\mu^2)^t} (q^{-\rho})
	\sum_\eta s_{\mu^1 /\eta} (q^{(\mu^2)^t + \rho}) s_{(\mu^3)^t /\eta} (q^{\mu^2 + \rho}),
\end{align}
where $\kappa_\mu=\sum_{i=1}^{l(\mu)} \mu_i(\mu_i -2i +1)$,
and $q^{\mu+\rho}$ is the sequence
\begin{equation*}
	q^{\mu+\rho} =
	( q^{\mu_1 - \half}, q^{\mu_2-\frac{3}{2}}, \cdots, q^{\mu_l - l+\half},
	q^{ -l-\half}, q^{-l-\frac{3}{2}},\cdots ).
\end{equation*}
The toric diagram of a general smooth toric Calabi--Yau threefold is always a trivalent planar graph.
Each vertex of this diagram corresponds to a $\mathbb{C}^3$ piece.
Then the gluing rules say that the open Gromov--Witten invariants of this Calabi--Yau threefold can be obtained by gluing all these $\mathbb{C}^3$ pieces.
In this paper,
we focus on the double-$\mathbb{P}^1$ model (see Subsection 5.2 in \cite{HSS}, see also \cite{IK06,S06}).
The toric diagram of the double-$\mathbb{P}^1$ is Figure \ref{eqn:toric of dP}.
We consider the double-$\mathbb{P}^1$ model with one nontrivial open sector.
Thus,
the open string amplitude of this double-$\mathbb{P}^1$ model with only one nontrivial representation labeled by $\mu$ can be obtained by the gluing rule (see \cite{AKMV,LLLZ}) as follows
\begin{align}\label{eqn:open sector}
	Z^{double-\mathbb{P}^1}_{\mu;t_1,t_2}
	=\sum_{\lambda_1,\lambda_2}
	(-Q_1)^{|\lambda_1|}(-Q_2)^{|\lambda_2|}
	\mathcal{W}_{\lambda_2,\mu,\lambda_1}
	\mathcal{W}_{\lambda_1^t,\emptyset,\emptyset}
	\mathcal{W}_{\lambda_2^t,\emptyset,\emptyset}.
\end{align}
A unified approach to simplify the above formula can be seen in \cite{IK06}.
They systematically studied the toric Calabi--Yau threefold, whose toric diagram can be represented as a dual graph of a strip, and the resulting formulas make vastly simplification.
For the full string amplitude,
one still needs to multiply the contribution from the closed sector to the above equation \eqref{eqn:open sector} (see, for example, equation (2.1) in \cite{S06}).

{\bf Proof of the Corollary \ref{cor:dP1}:}
The open string amplitude of double-$\mathbb{P}^1$ model with only one nontrivial representation $\mu$ is (see \cite{IK06}, see also equation (2.6) in \cite{TN16} for this special case of double-$\mathbb{P}^1$ model)
\begin{align}\label{eqn:Zmu from dP}
	Z^{double-\mathbb{P}^1}_{\mu;t_1,t_2}
	=
	\frac{s_{\mu^t}(q^{-\rho})
	\cdot \prod_{i,j=1}^\infty (1-Q_2q^{-\mu_j+i+j-1})
	\prod_{i,j=1}^\infty (1-Q_1q^{-\mu^t_i+i+j-1})}
	{\prod_{i,j=1}^\infty (1-Q_1Q_2q^{i+j-1})},
\end{align}
where the K{\"a}hler parameters $t_1, t_2$ are given by $Q_1=e^{-t_1}=q^L, Q_2=e^{-t_2}=q^N$.
For the denominator of above equation \eqref{eqn:Zmu from dP},
\begin{align*}
	\prod_{i,j=1}^\infty (1-Q_1Q_2q^{i+j-1})
	=\prod_{i,j=1}^\infty(1-q^{N+L+i+j-1})
\end{align*}
For the second factor in the numerator of the above equation \eqref{eqn:Zmu from dP},
for each fixed $j$,
we change $i$ to $i+\mu_j$,
and then obtain
\begin{align*}
	\prod_{i,j=1}^\infty (1-Q_2q^{-\mu_j+i+j-1})
	=\prod_{(i,j)\in\mu}(1-q^{N-c(i,j)})
	\cdot \prod_{i,j=1}^\infty(1-q^{N+i+j-1}).
\end{align*}
Similarly,
the third factor in the numerator of the above equation \eqref{eqn:Zmu from dP} has the following equality
\begin{align*}
	\prod_{i,j=1}^\infty (1-Q_1q^{-\mu^t_i+i+j-1})
	=\prod_{(i,j)\in\mu}(1-q^{L+c(i,j)})
	\cdot \prod_{i,j=1}^\infty(1-q^{L+i+j-1}).
\end{align*}
Combining all the above three equations,
and equation \eqref{eqn:schur q^rho},
we know that,
after multiplying a MacMahon factor $M(q)$,
the open string amplitude of double-$\mathbb{P}^1$ model $Z^{double-\mathbb{P}^1}_{\mu;t_1,t_2}$ is equal to
\begin{align}\label{eqn:Zmu final}
	q^{n(\mu^t)+|\mu|/2}\cdot
	\frac{\prod\limits_{(i,j)\in\mu} (1-q^{N-c(i,j)})
		(1-q^{L+c(i,j)})}
	{\prod\limits_{(i,j)\in\mu} (1-q^{h(i,j)})}
	\cdot \prod_{i,j=1}^\infty
	\frac{(1-q^{L+i+j-1})(1-q^{N+i+j-1})}
	{(1-q^{i+j-1})(1-q^{L+N+i+j-1})}.
\end{align}
It is obvious that,
the last factor in the above equation can be rewritten as
\begin{align*}
	\prod_{i,j=1}^\infty
	\frac{(1-q^{L+i+j-1})(1-q^{N+i+j-1})}
	{(1-q^{i+j-1})(1-q^{L+N+i+j-1})}
	=\prod_{1\leq n\leq N \atop 1\leq l\leq L}
	(1-q^{n+l-1})^{-1}.
\end{align*}
Thus, this corollary is proved by comparing equation \eqref{eqn:Zmu final} and Theorem \ref{thm:main ex}.
$\Box$

\section{The generating functions of certain plane partitions}
\label{sec:main}
In this section,
we give a precise study of the diagonal plane partitions and perpendicular plane partitions.
We will give formulas for their partition functions,
provide a simple example to distinguish their difference,
and offer a proof of their equivalence when all the $L,N,M$ go to infinity.

The first kind of object we study in this paper is the so-called diagonal plane partition (see \cite{ORV}),
which has diagonal boundaries along $x$ and $y$-axis directions and a perpendicular boundary along $z$-axis direction.
To be precise,
we first need to fix three finite positive integers $L, N, M$ and three partitions $\lambda, \mu, \nu$.
The diagonal plane partitions considered here are plane partitions contained in the region
\begin{align}\label{eqn:region}
	D(L,N,M)
	:=\{(x,y,z)\in\mathbb{R}^3_{\geq0}|x-y\leq L, y-x\leq M, z\leq N\}
\end{align}
as explained in Remark \ref{rmk:pi diagonal}.
By definition,
the set of diagonal plane partitions labeled by $(L, N, M)$ and $(\lambda, \mu, \nu)$ are
\begin{align}\label{eqn:def tP}
	\begin{split}
	\tilde{\mathcal{P}}^{L,N,M}_{\lambda,\mu,\nu}
	:=\{\pi=(\mu^k)_{k=-L}^{M}&|
	\pi_{i,j}=N\text{\ if\ }j\leq\mu^t_i,
	\pi_{i,j}<N\text{\ if\ }j>\mu^t_i,\text{\ and}\\
	&\mu^{-L+1}=\lambda^t, \mu^{-L}=\emptyset,
	\mu^{M-1}=\nu, \mu^M=\emptyset\}.
	\end{split}
\end{align}
Here, $\pi$ is a plane partition contained in the region $D(L,N,M)$,
which is equivalent to considering the following interlacing conditions for $(\mu^k)_{k=-L}^{M}$,
\begin{align*}
	\mu^{-L+1}\prec\mu^{-L+2}\prec\cdots
	\prec\mu^0\succ\cdots\succ\mu^{M-2}\succ\mu^{M-1}.
\end{align*}
Notice again that,
in the above definition,
even $\emptyset=\mu^{-L}\nprec\mu^{-L+1}=\lambda^t$ in general,
this does not cause any problem since $\pi=(\mu^k)_{k=-L}^{M}$ is regarded as a plane partition in the region $D(L,N,M)$ as Remark \ref{rmk:pi diagonal}.

Intuitively,
these conditions $\pi_{i,j}=N\text{\ if\ }j\leq\mu^t_i, \pi_{i,j}<N\text{\ if\ }j>\mu^t_i$
mean that the height of this plane partition is restricted to be weakly less than $N$, and the horizontal section at $z=N$ of this plane partition is just the Young diagram $\mu^t$.
These conditions $\mu^{-L+1}=\lambda^t, \mu^{-L}=\emptyset,
\mu^{M-1}=\nu, \mu^M=\emptyset$
mean that,
the diagonal vertical sections of this plane partition are $\lambda^t$ and $\mu$ at $x-y=L$ and $y-x=M$ planes respectively.
Outside the $x-y=L$ and $y-x=M$ planes,
there is no cube.
The following Figure \ref{eqn:ex 3D in tP} is a typical plane partition in $\tilde{\mathcal{P}}^{3,6,3}_{(1,1),(1,1),(4,2)}$.
It corresponds to the plane partition $\pi$ in equation \eqref{eqn:pi diagonal ex}.
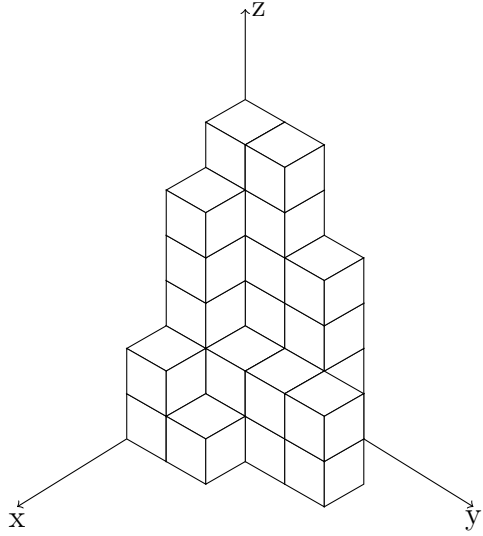
\begin{figure}[htbp]
	\begin{tikzpicture}[scale=0.5]
		\planepartition{{6,6,4},{5,2,2,2},{2,1}};
		\draw [->] (-2.5-0.1,-1.5)--(-5,-3);
		\node at(-5,-3.3) {x};
		\draw [->] (2.5+0.1,-1.5)--(5,-3);
		\node at(5,-3.3) {y};
		\draw [->] (0,6)--(0,8);
		\node at(0.3,8) {z};
	\end{tikzpicture}
	\caption{A 3D diagram in the region $D(3,6,3)$ and in $\tilde{\mathcal{P}}^{3,6,3}_{(1,1),(1,1),(4,2)}$.}
	\label{eqn:ex 3D in tP}
\end{figure}

Another interesting object we study is the perpendicular plane partition,
which has three perpendicular boundaries along $x,y$ and $z$-axis directions.
They are also introduced in \cite{ORV}.
However,
their method for calculating the partition function is essentially applied to the diagonal plane partition.
So we need to derive the correct formula for the perpendicular plane partition and clarify their difference,
which will be done in the next three subsections
and especially,
a simple example to distinguish them will be shown in Subsection \ref{sec:ex}.

The set of the perpendicular plane partitions is defined by
\begin{align*}
	\mathcal{P}^{L,N,M}_{\lambda,\mu,\nu}
	:=\{\pi=(\mu^k)_{k=-\infty}^{+\infty}&|
	\pi_{i,j}=N\text{\ if\ }j\leq\mu^t_i,
	\pi_{i,j}<N\text{\ if\ }j>\mu^t_i, \text{\ and}\\
	&\pi_{L,j}=\lambda^t_j, \pi_{L+1,j}=0,
	\pi_{j,M}=\nu_j, \pi_{j,M+1}=0, \forall j>0\}.
\end{align*}
Also,
intuitively,
these conditions $\pi_{i,j}=N\text{\ if\ }j\leq\mu^t_i, \pi_{i,j}<N\text{\ if\ }j>\mu^t_i$
mean that the height of this plane partition is restricted to be weakly less than $N$, and the horizontal section at $z=N$ of this plane partition is the Young diagram $\mu^t$.
These conditions $\pi_{L,j}=\lambda^t_j, \pi_{L+1,j}=0,
\pi_{j,M}=\nu_j, \pi_{j,M+1}=0, \forall j>0$
mean that,
the perpendicular section of this partition is $\lambda^t$ and $\mu$ at $x=L$ and $y=M$ planes respectively.
Outside the $x=L$ and $y=M$ planes,
there is no cube.
The following Figure \ref{eqn:ex 3D in P} is a plane partition in $\mathcal{P}^{3,6,4}_{(2,1,1),(1,1),(4,2)}$,
\begin{figure}[htbp]
	\includegraphics[scale=0.4]{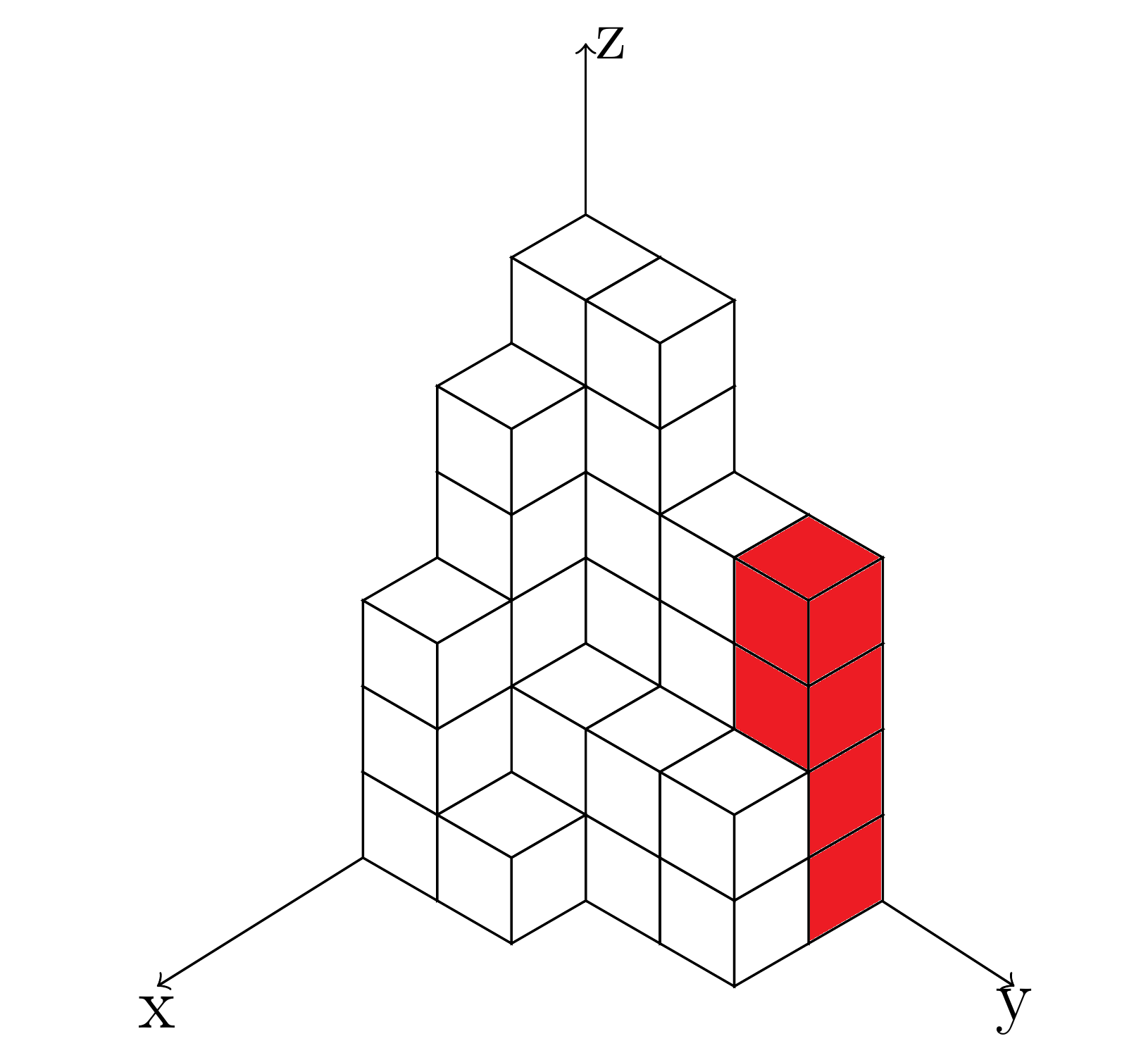}
%	\begin{tikzpicture}[scale=0.5]
%		\planepartition{{6,6,4,4},{5,2,2,2},{3,1}};
%		\draw [->] (-2.5-0.1,-1.5)--(-5,-3);
%		\node at(-5,-3.3) {x};
%		\draw [->] (3.5-0.05,-2)--(5,-3);
%		\node at(5,-3.3) {y};
%		\draw [->] (0,6)--(0,8);
%		\node at(0.3,8) {z};
%		\draw [->](2.7,5)--(2.7,2);
%	\node at(5,5.3){the different part};\node at(6,4.5){from Figure \ref{eqn:ex 3D in tP}};
%	\end{tikzpicture}
	\caption{A 3D diagram in $\mathcal{P}^{3,6,4}_{(2,1,1),(1,1),(4,2)}$.
	The red part is different from the example shown in Figure \ref{eqn:ex 3D in tP}.}
	\label{eqn:ex 3D in P}
\end{figure}

\indent Another example is $\pi$ in equation \ref{eqn:pi ex},
which could be considered as a plane partition in $\mathcal{P}^{3,6,3}_{(2),(1,1),(3,2)}$.

\subsection{Diagonal plane partitions}
\label{sec:diagonal boundary}
The method in this subsection mainly follows from \cite{ORV} dealing with the crystal melting model.

We are interested in the following partition function of plane partitions
\begin{align}\label{eqn:def tZ LNM}
	\tilde{Z}^{L,N,M}_{\lambda,\mu,\nu}
	=q^{-\binom{\lambda}{2}-\binom{\nu^t}{2}}
	\cdot \sum_{\pi\in\tilde{\mathcal{P}}^{L,N,M}_{\lambda,\mu,\nu}}
	q^{|\pi|}
	\in\mathbb{Z}[q,q^{-1}]
\end{align}
for finite positive integers $L,N,M$,
where $\binom{\lambda}{2}:=\sum_{i}\binom{\lambda_i}{2}$
and the extra factor $q^{-\binom{\lambda}{2}-\binom{\nu^t}{2}}$ appears just for matching up the notation in \cite{ORV} (see equations (3.15) and (3,17) in \cite{ORV}) and it will not cause any difficulty.
The above generating function contains much information.
For example,
the number of plane partitions in $\tilde{\mathcal{P}}^{L,N,M}_{\lambda,\mu,\nu}$ can be read by letting $q=1$ in $\tilde{Z}^{L,N,M}_{\lambda,\mu,\nu}$.
So we are interested in finding a formula to calculate the above partition function.

We will also study the case that some of $L,N,M$ go to infinity.
In that case,
unless corresponding $\lambda,\mu,\nu$ are empty partitions,
the size $|\pi|$ is infinite,
and then $q^{|\pi|}$ makes no sense since the power is infinite.
Thus, we will consider the following substitution (see equation (3.14) in \cite{ORV}).
Here, we take that all $L,N,M$ go to infinity as an example (one can similarly deal with the cases that only one or two of $L,N,M$ go to infinity).
In this case,
the corresponding partition function is defined by
\begin{align}\label{eqn:def tz inf}
	\tilde{Z}^{\infty,\infty,\infty}_{\lambda,\mu,\nu}:=
	\lim_{L,N,M\rightarrow\infty}
	q^{-L|\lambda|-N|\mu|-M|\nu|}
	\cdot\tilde{Z}^{L,N,M}_{\lambda,\mu,\nu}
	\in\mathbb{Z}[\![q]\!][q^{-1}].
\end{align}
The above limit exists (see equation (3.14) in \cite{ORV}).
Moreover, the above limit can also be directly regarded as an ordinary generating function weighted by a kind of modified size of infinite diagonal plane partitions.
For that,
we define
\begin{align*}
	\tilde{\mathcal{P}}^{\infty,\infty,\infty}_{\lambda,\mu,\nu}
	:=\{\pi=(\mu^k)_{k=-\infty}^{+\infty}&|
	\pi_{i,j}=\infty\text{\ if\ }j\leq\mu^t_i,
	\exists K>0 \text{\ such\ that\ } \pi_{i,j}<K\text{\ for\ all\ }j>\mu^t_i\\
	&\mu^{k}=\lambda^t \text{\ if\ }k\ll0,
	\mu^{k}=\nu \text{\ if\ }k\gg0\}.
\end{align*}
In this case,
the size of each plane partition in $\tilde{\mathcal{P}}^{\infty,\infty,\infty}_{\lambda,\mu,\nu}$ is infinite so long as one of $\lambda,\mu,\nu$ is not the empty partition.
So we need to define the modified size of plane partitions in $\tilde{\mathcal{P}}^{\infty,\infty,\infty}_{\lambda,\mu,\nu}$ to make sense of $q^{|\pi|}$ and the corresponding partition function.
For each $\pi\in\tilde{\mathcal{P}}^{\infty,\infty,\infty}_{\lambda,\mu,\nu}$,
there exists sufficient large integer $K>0$ such that
\begin{align}\label{eqn:K condition diagonal}
	\pi_{i,j}\leq K \text{\ for\ all\ }j>\mu^t_i,
	\mu^{k}=\lambda^t \text{\ if\ }k\leq -K,
	\mu^{k}=\nu \text{\ if\ }k\geq K.
\end{align}
We define $\pi^K$ as the plane partition obtained from $\pi$ by restricting it to the region
\begin{align*}
	D(K,K,K)=\{(x,y,z)\in\mathbb{R}^3_{\geq0}|x-y\leq K, y-x\leq K, z\leq K\}.
\end{align*}
And the condition \eqref{eqn:K condition diagonal} is equivalent to saying that $\pi^K$ already belongs to the set $\tilde{\mathcal{P}}^{K,K,K}_{\lambda,\mu,\nu}$.
Thus, the modified size of infinite diagonal plane partition $\pi$ is defined by
\begin{align*}
	\widetilde{|\pi|}:
	=|\pi^K|-K\cdot(|\lambda|+|\mu|+|\nu|).
\end{align*}
It is obvious that the modified size of $\pi$ is well-defined and independent of the choice of sufficient large $K$
since the different reasonable choices of $\pi^K$ and $\pi^{K'}$ only differ by some cubes whose number is exactly $|K-K'|\cdot(|\lambda|+|\mu|+|\nu|)$.
Also, when $\lambda=\mu=\nu=\emptyset$,
the modified size is equal to the standard size.
As a consequence,
we have 
\begin{align}\label{eqn:def tZ infty}
	\tilde{Z}^{\infty,\infty,\infty}_{\lambda,\mu,\nu}
	=q^{-\binom{\lambda}{2}-\binom{\nu^t}{2}}
	\cdot \sum_{\pi\in\tilde{\mathcal{P}}^{\infty,\infty,\infty}_{\lambda,\mu,\nu}}
	q^{\widetilde{|\pi|}}.
\end{align}
That is to say,
$\tilde{Z}^{\infty,\infty,\infty}_{\lambda,\mu,\nu}$
can also be considered as the partition function of plane partitions without finiteness.
One can similarly deal with the cases when only one or two of $L,N,M$ go to infinity.

When $L=N=M=\infty$ and $\lambda=\mu=\nu=\emptyset$,
that is to say,
there is no any restriction on such plane partitions.
The above generating function is thus a summation over all the finite plane partitions,
which corresponds to the unbounded case of the MacMahon formula,
i.e. equation \eqref{eqn:macmahon} by letting $a,b,c\rightarrow\infty$.
One of the interesting things is that
the MacMahon formula is a product of some simple rational functions of $q$.
Thus,
it will be interesting to see whether the general cases also have product formulas.
Throughout the rest of this subsection,
we will prove the equation \eqref{eqn:main tZ} in Theorem \ref{thm:main},
which gives a formula for $\tilde{Z}^{L,N,M}_{\lambda,\mu,\nu}$ in terms of vacuum expectation value.

The main method of this subsection was originally used in \cite{ORV}.
They match the partition functions to matrix elements of product of vertex operators.
Their method is applied to the $N=\infty$ case.
We generalize it to general $N<\infty$ by inserting certain projection operators.
Moreover,
we clarify that
their method is essentially used to compute the partition function of diagonal plane partitions
(even though they said that their method is for perpendicular plane partitions (see Subsection 3.4 in \cite{ORV})).
We first introduce the notations in equation \eqref{eqn:main tZ}.

For a partition $\mu=(m_1,...,m_r|n_1,...,n_r)$,
denote by the modified vertex operators as
\begin{align}
	\Gamma^{d}_{+,\{j,\mu\}}(z)
	=\begin{cases}
		\mathbbm{1}_{l(\cdot^t)\leq d}
		\cdot \Gamma_+(z), &\text{\ if\ }j\notin\{m_1,m_2,...,m_r\},\\
		\mathbbm{1}_{l(\cdot^t)\leq d}
		\cdot \Gamma_-(z^{-1}), &\text{\ if\ }j\in\{m_1,m_2,...,m_r\}
	\end{cases}
\end{align}
and
\begin{align}
	\Gamma^{d}_{-,\{i,\mu\}}(z)
	=\begin{cases}
		\Gamma_-(z)
		\cdot \mathbbm{1}_{l(\cdot^t)\leq d}, &\text{\ if\ }i\notin\{n_1,n_2,...,n_r\},\\
		\Gamma_+(z^{-1})
		\cdot \mathbbm{1}_{l(\cdot^t)\leq d}, &\text{\ if\ }i\in\{n_1,n_2,...,n_r\}
	\end{cases}.
\end{align}
The $\mathbbm{1}_{l(\cdot)\leq d}$ is the operator that projects onto the subspace spanned by $|\mu\rangle$ with $l(\mu)\leq d$.
Similarly, the operator $\mathbbm{1}_{l(\cdot^t)\leq d}$ projects onto the subspace spanned by $|\mu\rangle$ with $l(\mu^t)\leq d$,
which is equivalent to the condition $\mu_1\leq d$.
When $d=\infty$, $\mathbbm{1}_{l(\cdot)\leq \infty}$
and $\mathbbm{1}_{l(\cdot^t)\leq \infty}$ are the identity operator.

\begin{thm}[= equation \eqref{eqn:main tZ} in Theorem \ref{thm:main}]
	\label{thm:sZ}
	The partition function of diagonal plane partitions has the following formula
	\begin{align}\label{eqn:tZ vev}
		\begin{aligned}
		\tilde{Z}^{L,N,M}_{\lambda,\mu,\nu}
		=&\delta_{L>\mu_1}\delta_{M>\mu^t_1}
		\cdot q^{L|\lambda|+N|\mu|+M|\nu|}
		\cdot q^{-|\lambda|/2-|\nu|/2
			-\binom{\lambda}{2}-\binom{\nu^t}{2}}\\
		&\cdot\langle \lambda^t|
		\prod_{0\leq j<L-1}^{\longleftarrow} \Gamma^{N-1}_{+,\{j,\mu\}}(q^{j+\half})
		\cdot \mathbbm{1}_{l(\cdot^t)\leq N-1}
		\cdot \prod_{0\leq i<M-1}^{\longrightarrow} \Gamma^{N-1}_{-,\{i,\mu\}}(q^{i+\half})
		|\nu\rangle,
		\end{aligned}
	\end{align}
	where
	\begin{align*}
		\prod_{0\leq j<L-1}^{\longleftarrow} \Gamma^{N-1}_{+,\{j,\mu\}}(q^{j+\half})
		=\Gamma^{N-1}_{+,\{L-2,\mu\}}(q^{L-\frac{3}{2}})
		\cdots \Gamma^{N-1}_{+,\{0,\mu\}}(q^{\half})
	\end{align*}
	and
	\begin{align*}
		\prod_{0\leq i<M-1}^{\longrightarrow} \Gamma_{-,\{i,\mu\}}^{N-1}(q^{i+\half})
		=\Gamma_{-,\{0,\mu\}}^{N-1}(q^{\half})
		\cdots \Gamma_{-,\{M-2,\mu\}}^{N-1}(q^{M-\frac{3}{2}})
	\end{align*}
	Here the $L,N,M$ could be any positive integers or infinities,
	and $q^{L|\lambda|}, q^{N|\lambda|}, q^{M|\lambda|}$ should be understood as $1$ when corresponding $L, N, M=\infty$.
\end{thm}
{\bf Proof:}
At first,
if $L\leq\mu_1$ or $M\leq\mu_1^t$,
the set $\tilde{\mathcal{P}}^{L,N,M}_{\lambda,\mu,\nu}$ is empty by definition.
Thus, equation \eqref{eqn:tZ vev} automatically holds.
Moreover,
when some of $L, N, M$ go to infinity,
the right hand side of equation \eqref{eqn:tZ vev} is well-defined as an element in $\mathbb{Z}[\![q]\!][q^{-1}]$
and is compatible with the definition \eqref{eqn:def tz inf}.
Thus,
from now on,
we assume the condition $\infty>L\geq\mu_1, \infty>M\geq\mu_1^t$ and ignore the term $\delta_{L\geq\mu_1}\delta_{M\geq\mu^t_1}$ in equation \eqref{eqn:tZ vev}.

We first review the method used by Okounkov--Reshetikhin--Vafa \cite{ORV} for a special case.
When $N=\infty$ and $\mu=\emptyset$,
there is a one-to-one correspondence between plane partitions in the set $\tilde{\mathcal{P}}^{L,\infty,M}_{\lambda,\emptyset,\nu}$
and interlacing Young diagrams $(\mu^k)_{k=-L}^{M}$ satisfying
\begin{align}\label{eqn:mu=0 interlacing}
	\mu^{-L}=\emptyset,
	\mu^{-L+1}=\lambda^t
	\quad\text{and}\quad
	\mu^{M-1}=\nu,
	\mu^M=\emptyset,
\end{align}
where interlacing condition means
\begin{align*}
	\mu^{-L+1}\prec\mu^{-L+2}\prec\cdots
	\prec\mu^0\succ\cdots\succ\mu^{M-2}\succ\mu^{M-1}.
\end{align*}
Thus,
by definition,
the partition function of such plane partitions is equal to
\begin{align*}
	\tilde{Z}^{L,\infty,M}_{\lambda,\emptyset,\nu}
	=q^{-\binom{\lambda}{2}-\binom{\nu^t}{2}}
	\sum_{(\mu^k)_{k=-L}^{M}} q^{\sum_{k}|\mu^k|}
	=q^{-\binom{\lambda}{2}-\binom{\nu^t}{2}}
	\langle \lambda^t|
	\prod_{k=0}^{L-2} q^{L_0}\Gamma_+(1)
	\cdot q^{L_0}
	\cdot \prod_{k=0}^{M-2} \Gamma_-(1)q^{L_0}
	|\nu\rangle,
\end{align*}
which is the $\mu=\emptyset$ case of equation (3.17) in \cite{ORV}.
By splitting the middle $q^{L_0}$ in half and commuting all $L_0$ to outside,
we obtain
\begin{align*}
	\tilde{Z}^{L,\infty,M}_{\lambda,\emptyset,\nu}
	=q^{(L-1/2)|\lambda|+(M-1/2)|\nu|-\binom{\lambda}{2}-\binom{\nu^t}{2}}
	\cdot \langle \lambda^t|
	\prod_{k=0}^{L-2} \Gamma_+(q^{k+1/2})
	\cdot \prod_{k=0}^{M-2} \Gamma_-(q^{k+1/2})
	|\nu\rangle.
\end{align*}

About the $\mu\neq\emptyset$ case,
the brilliant method used by Okounkov--Reshetikhin--Vafa \cite{ORV} (see also \cite{OR07}) is that
the $\mu$-condition is equivalent to considering the so-called skew plane partition.
More precisely,
for a partition $\pi\in\tilde{\mathcal{P}}^{L,N,M}_{\lambda,\mu,\nu}$,
now that for each $(i,j)$ such $j\leq\mu_i^t$ the $\pi_{i,j}$ must be $N$,
we can ignore all these $N\times |\mu|$ cubes which are supported on some $(i,j)$ such $j\leq\mu_i^t$.
As a result, we can associate $\pi$ a unique skew plane partition inside a container whose support is
(see Figure 5 in \cite{ORV})
\begin{align*}
	[0,L]\times[0,M]\setminus\mu.
\end{align*}
When $N=\infty$,
the result was obtained by Okounkov--Reshetikhin--Vafa as equation (3.17) in \cite{ORV}.
Next,
we directly generalize their method to $N<\infty$ case.
For any plane partition $\pi$ in the set $\tilde{\mathcal{P}}^{L,N,M}_{\lambda,\mu,\nu}$,
one can delete the cubes in the region $\{(x,y,z)|y\leq \mu^t_{\lfloor x\rfloor+1}\text{\ for\ } 0\leq x<L\}$
and thus transfer $\pi$ to be a skew plane partition $\pi'$ in the sense of \cite{OR07}.
As a consequence,
the skew plane partition $\pi'$ one-to-one corresponds to a series of Young diagrams $(\mu^k)_{k=-L}^{M}$ satisfying the conditions \eqref{eqn:mu=0 interlacing},
$\mu^k_1\leq N-1$ for all $k$ and
\begin{align*}
	\mu^{j-1}\ (\prec)_{j,\mu}\ \mu^{j}, -L+2<j\leq0,
	\quad \mu^{i}\ (\succ)_{i,\mu}\ \mu^{j+1}, 0\leq i<M-2,
\end{align*}
where $(\prec)_{j,\mu}=\prec$ if $-j\notin\{m_1,...,m_r\}$, otherwise it is $\succ$,
and similarly,
$(\succ)_{i,\mu}=\succ$ if $i\notin\{n_1,...,n_r\}$, otherwise it is $\prec$.
By the definition of our modified vertex operators $\Gamma^{N-1}_{\pm,\{j,\mu\}}(q)$,
their actions exactly generate a series of such Young diagrams.
Thus,
since the difference between the sizes of $\pi$ and $\pi'$ is $N|\mu|$,
we have
\begin{align*}
	&\tilde{Z}^{L,N,M}_{\lambda,\mu,\nu}
	=q^{N|\mu|-\binom{\lambda}{2}-\binom{\nu^t}{2}}
	\cdot \sum_{(\mu^k)_{k=-L}^{M}\text{\ satisfying\ certain\ conditions}} q^{\sum_{k}|\mu^k|}\\
	&\quad\ =q^{N|\mu|-\binom{\lambda}{2}-\binom{\nu^t}{2}}
	\cdot \langle \lambda^t|
	\prod_{0\leq j<L-1}^{\longleftarrow} q^{L_0}\Gamma^{N-1}_{+,\{j,\mu\}}(1)
	\cdot q^{L_0}
	\cdot \mathbbm{1}_{l(\cdot^t)\leq N-1}
	\cdot \prod_{0\leq i<M-1}^{\longrightarrow} \Gamma^{N-1}_{-,\{i,\mu\}}(1) q^{L_0}
	|\nu\rangle.
\end{align*}
As a consequence,
by moving all $q^{L_0}$ to outside,
equation \eqref{eqn:tZ vev} is obtained.
$\Box$

\begin{rmk}\label{rmk:empty = wall}
	When $\lambda=\emptyset$ or $\nu=\emptyset$,
	the diagonal condition with boundary $\emptyset$ is equivalent to adding a wall at $x=L$ or $y=M$ respectively,
	and the plane partitions are bounded by the corresponding wall
	(see, for example, the explanation in Section 2 in \cite{Oku05}).
	The reason is that,
	by the interlacing conditions,
	the plane partition has a empty boundary in the slice $x=y\pm L$ will enforce the slice at $x=y\pm (L-1)$ has at most one row,
	the slice at $x=y\pm (L-2)$ has at most two rows,
	etc.
	These conditions are equivalent to putting a wall at $x=y\pm L$.
\end{rmk}

\subsection{Perpendicular plane partitions}

The partition function of the perpendicular plane partitions labeled by $(L,N,M)$ and $(\lambda,\mu,\nu)$ is
\begin{align*}
	Z^{L,N,M}_{\lambda,\mu,\nu}
	:=\sum_{\pi\in\mathcal{P}^{L,N,M}_{\lambda,\mu,\nu}}
	q^{|\pi|}
	\in\mathbb{Z}[q]
\end{align*}
for finite positive integers $L,N,M$.
We are also interested in the $L,N,M\rightarrow\infty$ case,
and the corresponding partition function is defined by
\begin{align}\label{eqn:def z inf}
	Z^{\infty,\infty,\infty}_{\lambda,\mu,\nu}:=
	\lim_{L,N,M\rightarrow\infty}
	q^{-L|\lambda|-N|\mu|-M|\nu|}
	Z^{L,N,M}_{\lambda,\mu,\nu}
	\in\mathbb{Z}[\![q]\!][q^{-1}].
\end{align}
Similar to the case in Subsection \ref{sec:diagonal boundary},
the above limit of generating functions can also be regarded as an ordinary generating function weighted by a new kind of modified size of infinite perpendicular plane partitions.
The modified size $\overline{|\pi|}$ of the perpendicular type is defined as follows.
First,
the set of infinite perpendicular plane partitions is defined as
\begin{align*}
	\mathcal{P}^{\infty,\infty,\infty}_{\lambda,\mu,\nu}
	=&\{\pi=(\mu^k)_{k=-\infty}^{+\infty}|
	\pi_{i,j}=\infty\text{\ if\ }j\leq\mu^t_i,
	\exists K>0 \text{\ such\ that\ }\pi_{i,j}<K\text{\ for\ all\ }j>\mu^t_i,\\
	&\text{\ and\ for\ } L, M\gg0,
	\pi_{L,j}=\lambda^t_j, \pi_{L+1,j}=0,
	\pi_{j,M}=\nu_j, \pi_{j,M+1}=0, \forall j>0\}
\end{align*}
Thus, for each $\pi\in\mathcal{P}^{\infty,\infty,\infty}_{\lambda,\mu,\nu}$,
there exists a sufficient large $K>0$ such that,
when denote by $\pi^K$ the restriction of $\pi$ in the region
$[0,K]\times[0,K]\times[0,K]$,
$\pi^K$ belongs to the set $\mathcal{P}^{K,K,K}_{\lambda,\mu,\nu}$.
Thus, the modified size of infinite perpendicular plane partition $\pi$ is defined by
\begin{align*}
	\overline{|\pi|}:=|\pi^K|-K\cdot(|\lambda|+|\mu|+|\nu|).
\end{align*}
Similar to the Subsection \ref{sec:diagonal boundary},
one can see that this kind of modified size is well-defined.
As a consequence,
\begin{align*}
	Z^{\infty,\infty,\infty}_{\lambda,\mu,\nu}=
	\sum_{\pi\in\mathcal{P}^{\infty,\infty,\infty}_{\lambda,\mu,\nu}}
	q^{\overline{|\pi|}}.
\end{align*}
One can similarly deal with the cases when only one or two of $L,N,M$ go to infinity.

For obtaining formula for the partition function $Z^{L,N,M}_{\lambda,\mu,\nu}$ of perpendicular plane partitions,
we need to introduce the following new notations.
For a partition $\mu$ whose Frobenius notation is $(m_1,...,m_r|n_1,...,n_r)$,
the modified vertex operators are
\begin{align}
	\Gamma^{d,(k,l)}_{+,\{j,\mu\}}(z)
	=\begin{cases}
		\mathbbm{1}_{l(\cdot^t)\leq d}
		\cdot \Gamma_+(z)
		\cdot \mathbbm{1}_{(k,l)},
		&\text{\ if\ }j\notin\{m_1,m_2,...,m_r\},\\
		\mathbbm{1}_{l(\cdot^t)\leq d}
		\cdot \Gamma_-(z^{-1})
		\cdot\mathbbm{1}_{(k,l)},
		&\text{\ if\ }j\in\{m_1,m_2,...,m_r\},
	\end{cases}
\end{align}
and
\begin{align}
	\Gamma^{d,(k,l)}_{-,\{j,\mu\}}(z)
	=\begin{cases}
		\mathbbm{1}_{(k,l)}
		\cdot \Gamma_-(z)
		\cdot \mathbbm{1}_{l(\cdot^t)\leq d},
		&\text{\ if\ }j\notin\{n_1,n_2,...,n_r\},\\
		\mathbbm{1}_{(k,l)}
		\cdot \Gamma_+(z^{-1})
		\cdot \mathbbm{1}_{l(\cdot^t)\leq d},
		&\text{\ if\ }j\in\{n_1,n_2,...,n_r\},
	\end{cases}
\end{align}
where $\mathbbm{1}_{l(\cdot^t)\leq d}$ is the projection operator defined in Subsection \ref{sec:diagonal boundary}
and $\mathbbm{1}_{(k,l)}$ is the operator projects onto the subspace
\begin{align*}
	\span\{|\mu\rangle\ |\ \mu_{k}=l, \mu_{k'}=0\text{\ for\ }k'>k \}
\end{align*}
when $k>0$,
and $\mathbbm{1}_{(k,l)}$ is the identity when $k\leq0$.

\begin{thm}[= equation \eqref{eqn:main Z} in Theorem \ref{thm:main}]
	The partition function of perpendicular plane partitions has the following formula,
	for any positive integers $L,N,M$,
	\begin{align}\label{eqn:Z vev}
		\begin{split}
		&Z^{L,N,M}_{\lambda,\mu,\nu}
		=\delta_{L>\mu_1}\delta_{M>\mu^t_1}
		\cdot \tilde{\delta}_{L,N,\mu,\lambda^t}
		\tilde{\delta}_{M,N,\mu^t,\nu}
		\cdot q^{L\lambda^t_1+N|\mu|+M\nu_1}
		\cdot q^{-\lambda^t_1/2-\nu_1/2}\\
		&\quad\ \cdot\langle (\lambda^t_1)|
		\prod_{0\leq j<L-1}^{\longleftarrow} \Gamma^{N-1,(k'_j,\lambda^t_{L-j})}_{+,\{j,\mu\}}(q^{j+\half})
		\cdot \mathbbm{1}_{l(\cdot^t)\leq N-1}
		\cdot \prod_{0\leq i<M-1}^{\longrightarrow} \Gamma^{N-1,(k_i,\nu_{M-i})}_{-,\{i,\mu\}}(q^{i+\half})
		|(\nu_1)\rangle,
		\end{split}
	\end{align}
	where $k_i=M-i-\#\{k|\mu^t_{k}>i+1\},
	k'_j=L-j-\#\{k|\mu_{k}>j+1\}$,
	\begin{align*}
		\tilde{\delta}_{L,N,\mu,\lambda^t}=
		\begin{cases}
			1, &\text{\ if\ } \lambda^t_{i}=N \text{\ for\ all\ }1\leq i\leq\#\{k|\mu_k=L\},\\
			0, &\text{\ otherwise},
		\end{cases}
	\end{align*}
	and $\tilde{\delta}_{M,N,\mu^t,\nu}$ is defined similarly.
\end{thm}
{\bf Proof:}
The method is similar to the proof of Theorem \ref{thm:sZ}.
First,
the appearance of the term $\tilde{\delta}_{L,N,\mu,\lambda^t}
\tilde{\delta}_{M,N,\mu^t,\nu}$ is equivalent to requiring that,
if $\mu_1=L$,
then $\lambda^t_1$ must be $N$ since in this case $\pi_{L,1}=N$,
if $\mu_2=L$,
then $\lambda^t_2$ must be $N$ since in this case $\pi_{L,2}=N$,
and so on.
Thus,
for convenience,
we can ignore the terms $\delta_{L\geq\mu_1}\delta_{M\geq\mu^t_1}
\cdot \tilde{\delta}_{L,N,\mu,\lambda^t}
\tilde{\delta}_{M,N,\mu^t,\nu}$.

For any plane partition $\pi$ in the set $\mathcal{P}^{L,N,M}_{\lambda,\mu,\nu}$,
by deleting the cubes in the region $\{(x,y,z)|y\leq \mu^t_{\lfloor x\rfloor+1}\text{\ for\ } 0\leq x<L\}$,
one thus transfer $\pi$ to be a skew plane partition $\pi'$ in the sense of \cite{OR07}.
By dividing $\pi'$ along the diagonal slices,
there is a one-to-one correspondence between it and a series of interlacing Young diagrams $(\mu^k)_{k=-L}^{M}$ satisfying the following conditions:

i) $\mu^k_1\leq N-1$ for all $k$,

ii) $\mu^{j-1}\ (\prec)_{j,\mu}\ \mu^{j}$ for $-L+2<j\leq0$,
and $\mu^{i}\ (\succ)_{i,\mu}\ \mu^{i+1}$ for $0\leq i<M-2$,

iii) For $0\leq j<L-1$, $\mu^{-j}_{k'_j}=\lambda^t_{L-j}, \mu_{k'_j+1}^{-j}=0$,
and for $0\leq i<M-1$,
$\mu^{i}_{k_i}=\nu_{M-i},
\mu^{i}_{k_i+1}=0$.\\
The third condition above is equivalent to saying that
$\pi_{L,j}=\lambda^t_j$ and $\pi_{i,M}=\nu_i$,
which are the perpendicular boundary conditions of $\pi$ along $x$ and $y$-axis directions.
By the definition of our modified vertex operators $\Gamma^{N-1,(k,l)}_{\pm,\{j,\mu\}}(q)$,
their actions exactly generate a series of such Young diagrams.
Thus,
since the difference between the sizes of $\pi$ and $\pi'$ is $N|\mu|$,
we have
\begin{align*}
	\tilde{Z}^{L,N,M}_{\lambda,\mu,\nu}
	=&q^{N|\mu|}
	\cdot \sum_{(\mu^k)_{k=-L}^{M}\text{\ satisfying\ certain\ conditions}} q^{\sum_{k}|\mu^k|}\\
	=&q^{N|\mu|}
	\cdot \langle (\lambda^t_1)|
	\prod_{0\leq j<L-1}^{\longleftarrow} q^{L_0}\Gamma^{N-1,(k'_j,\lambda^t_{L-j})}_{+,\{j,\mu\}}(1)\\
	&\quad\quad\cdot q^{L_0}
	\cdot \mathbbm{1}_{l(\cdot^t)\leq N-1}
	\cdot \prod_{0\leq i<M-1}^{\longrightarrow} \Gamma^{N-1,(k_i,\nu_{M-i})}_{-,\{i,\mu\}}(1) q^{L_0}
	|(\nu_1)\rangle.
\end{align*}
Finally,
by moving all $q^{L_0}$ to outside,
equation \eqref{eqn:Z vev} is obtained.
$\Box$

\subsection{An example distinguishing two types of plane partitions}
\label{sec:ex}
In this subsection,
we give an example to clarify the difference between diagonal plane partitions and perpendicular plane partitions.

\begin{ex}
	We consider the following case.
	When $L=M=2, N=\infty$ and $\lambda=\nu=(1), \mu=\emptyset$.
	The following are all diagonal plane partitions in $\tilde{\mathcal{P}}^{2,\infty,2}_{(1),\emptyset,(1)}$,
	\begin{align}\label{eqn:ex dp}
		\left(
		\begin{matrix}
			k & 1 & \cdots\\
			1 & 0 & \cdots\\
			\vdots & \vdots & \ddots\\
		\end{matrix}\right), k\geq1,
	\quad\quad
	\left(
	\begin{matrix}
		k & 1 & \cdots\\
		1 & 1 & \cdots\\
		\vdots & \vdots & \ddots\\
	\end{matrix}\right), k\geq1.
	\end{align}
	Thus,
	the partition function of diagonal plane partitions in this case is
	\begin{align}\label{eqn:ex dp d}
		\tilde{Z}^{2,\infty,2}_{(1),\emptyset,(1)}
		=q^3+2\sum_{k\geq4}q^k
		=\frac{q^3(1+q)}{1-q}.
	\end{align}
	On the other hand,
	one can see that those plane partitions in the right hand side of equation \eqref{eqn:ex dp} do not belong to $\mathcal{P}^{2,\infty,2}_{(1),\emptyset,(1)}$.
	Thus,
	the partition function of perpendicular plane partitions in this case is just
	\begin{align}\label{eqn:ex dp p}
		Z^{2,\infty,2}_{(1),\emptyset,(1)}
		=\sum_{k\geq3}q^k
		=\frac{q^3}{1-q}.
	\end{align}
	Equations \eqref{eqn:ex dp d} and \eqref{eqn:ex dp p} can also be obtained from our formulas \eqref{eqn:main tZ} and \eqref{eqn:main Z} respectively.
	However,
	their difference is a multiplication factor $(1+q)$,
	but not $q^{\binom{\lambda}{2}+\binom{\nu^t}{2}}$ explained in equation (3.15) in \cite{ORV}.
\end{ex}

It seems that the difference between $\tilde{Z}^{L,N,M}_{\lambda,\mu,\nu}$
and
$Z^{L,N,M}_{\lambda,\mu,\nu}$
is very complicated in general cases.
It will be interesting to give a consistent answer.

\subsection{The equivalence of two partition functions when $L,N,M$ go to infinity}
\label{sec:equivalence}

\begin{prop}\label{prop:equivalence}
	When $L,N,M$ go to infinity,
	\begin{align*}
		\tilde{Z}^{\infty,\infty,\infty}_{\lambda,\mu,\nu}
		=Z^{\infty,\infty,\infty}_{\lambda,\mu,\nu}
		\in\mathbb{Z}[\![q]\!][q^{-1}].
	\end{align*}
\end{prop}
{\bf Proof:}
First,
we notice that if $\pi$ is a plane partition in $\mathcal{P}^{\infty,\infty,\infty}_{\lambda,\mu,\nu}$,
then for a sufficient large $K>0$ such that the boundaries of $\pi$ in the perpendicular region $[0,K]\times[0,K]\times[0,K]$ are described by $\lambda, \mu$ and $\nu$,
we must have the $m$-th diagonal slice of $\pi$ is $\nu$ when $m>K$ and is $\lambda$ if $m<-K$.
Thus,
$\pi$ is also a plane partition in the set $\tilde{\mathcal{P}}^{\infty,\infty,\infty}_{\lambda,\mu,\nu}$.
Conversely,
if we assume $\pi$ is a plane partition in $\tilde{\mathcal{P}}^{\infty,\infty,\infty}_{\lambda,\mu,\nu}$
such that the boundaries of $\pi$ in the diagonal region
\begin{align*}
	D(K,K,K)=\{(x,y,z)\in\mathbb{R}^3_{\geq0}|x-y\leq K, y-x\leq K, z\leq K\}
\end{align*}
are described by $\lambda,\mu$ and $\nu$,
then for any $m>K+|\lambda|+|\nu|$,
the boundaries of $\pi$ in the perpendicular region $[0,m]\times[0,m]\times[0,m]$ are also $\lambda,\mu$ and $\nu$.
Thus,
$\pi$ is also a plane partition in $\mathcal{P}^{\infty,\infty,\infty}_{\lambda,\mu,\nu}$.
As a consequence,
the two sets of plane partitions $\tilde{\mathcal{P}}^{\infty,\infty,\infty}_{\lambda,\mu,\nu}$ and
$\mathcal{P}^{\infty,\infty,\infty}_{\lambda,\mu,\nu}$
are the same.

By the definition of the modified sizes of diagonal type $\widetilde{|\pi|}$ and of perpendicular type $\overline{|\pi|}$,
their difference is
\begin{align*}
	\widetilde{|\pi|}
	=\overline{|\pi|}
	+\binom{\lambda}{2}+\binom{\nu^t}{2}
\end{align*}
since there exists a sufficient large $K$ such that both of $\widetilde{|\pi|}$ and $\overline{|\pi|}$ can be defined suitably when restricting $\pi$ in the region $D(K,K,K)$
and restricting $\pi$ in the region $[0,K]\times[0,K]\times[0,K]$ respectively,
and the number of cubes in the difference of these regions is \[\sum_{i}\sum_{j=1}^{\lambda_i}(j-1)
+\sum_{i}\sum_{j=1}^{\nu^t_i}(j-1)
=\binom{\lambda}{2}+\binom{\nu^t}{2}.\]
The term $q^{\binom{\lambda}{2}+\binom{\nu^t}{2}}$ is already considered in the definition of $\tilde{Z}^{\infty,\infty,\infty}_{\lambda,\mu,\nu}$ (see equation \eqref{eqn:def tZ infty},
see also equation (3.15) in \cite{ORV}).
Thus,
this proposition is proved.
$\Box$

\begin{cor}\label{cor:rotation}
	We have
	\begin{align*}
		\tilde{Z}^{L+1,N+1,\infty}_{\emptyset,\emptyset,\mu}
		=\tilde{Z}^{N+1,\infty,L+1}_{\emptyset,\mu,\emptyset}.
	\end{align*}
\end{cor}
{\bf Proof:}
The elements in the set $\tilde{\mathcal{P}}^{L+1,N+1,\infty}_{\emptyset,\emptyset,\mu}$
could be regarded as plane partitions restricted by two walls at $x=L+1$ and $z=N+1$ planes,
and having a diagonal limit boundary described by $\mu$ along the $y$-axis directions.
In terms of the proof in Proposition \ref{prop:equivalence},
the diagonal limit boundary condition is also equivalent to a perpendicular limit boundary condition.
Thus,
all the boundary conditions of those plane partitions in $\tilde{\mathcal{P}}^{L+1,N+1,\infty}_{\emptyset,\emptyset,\mu}$ along three directions can be regarded as perpendicular types,
and thus possess the rotation symmetry.
As a result,
by rotation,
elements in these two sets $\tilde{\mathcal{P}}^{L+1,N+1,\infty}_{\emptyset,\emptyset,\mu}$
and
$\tilde{\mathcal{P}}^{N+1,\infty,L+1}_{\emptyset,,\mu,\emptyset}$
have a one-to-one correspondence with each other. Thus, their partition functions are equal.
$\Box$

\section{Plane partitions with a limit boundary along $z$-axis direction}
\label{sec:case1}
In this section,
we study the plane partitions whose two directions are restricted by walls, and the other direction has a limit boundary.
We obtain a product formula for their partition function,
which is equivalent to the open string amplitude of double-$\mathbb{P}^1$ model with one nontrivial representation
(up to simple factors).
As a corollary,
we give a new derivation of the full MacMahon formula.

\subsection{Plane partitions bounded by two walls and possessing a limit boundary}

The plane partition $\pi$ in the set $\tilde{\mathcal{P}}^{N+1,\infty,L+1}_{\emptyset,\mu,\emptyset}$
satisfies that,
$\pi$ is bounded by two walls at $x=N+1$ and $y=L+1$ planes,
and the limit boundary of $\pi$ along the $z$-axis direction is described by $\mu$.
In this subsection,
we prove Theorem \ref{thm:main ex},
which gives a product formula for the corresponding partition function.
\begin{thm}[= Theorem \ref{thm:main ex}]
The partition function $\tilde{Z}^{N,\infty,L}_{\emptyset,\mu,\emptyset}$ has the following product formula
\begin{align}\label{eqn:Z^(N,infty,L)}
	\tilde{Z}^{N+1,\infty,L+1}_{\emptyset,\mu,\emptyset}
	=\delta_{N\geq\mu_1} \delta_{L\geq\mu^t_1}
	\cdot\prod_{1\leq n\leq N \atop 1\leq l\leq L}
	(1-q^{n+l-1})^{-1}
	\cdot
	\frac{\prod_{(i,j)\in\mu} (1-q^{N-c(i,j)})
	(1-q^{L+c(i,j)})}
	{\prod_{(i,j)\in\mu} (1-q^{h(i,j)})}.
\end{align}
\end{thm}
{\bf Proof:}
First, if $N<\mu_1$ or $L<\mu^t_1$,
by definition,
the set $\tilde{\mathcal{P}}^{N+1,\infty,L+1}_{\emptyset,\mu,\emptyset}$ is empty,
thus $\tilde{Z}^{N+1,\infty,L+1}_{\emptyset,\mu,\emptyset}=0$ trivially follows.
So we could always assume $N\geq\mu_1$, $L\geq\mu^t_1$ and ignore the $\delta_{N\geq\mu_1} \delta_{L\geq\mu^t_1}$ term, to prove this theorem.

By equation \eqref{eqn:main tZ},
the partition function $\tilde{Z}^{N+1,\infty,L+1}_{\emptyset,\mu,\emptyset}$ can be explicitly calculated in terms of the following vacuum expectation value
\begin{align}\label{eqn:z limit = vev}
	\tilde{Z}^{N+1,\infty,L+1}_{\emptyset,\mu,\emptyset}
	=\langle \emptyset|
	\prod_{0\leq j<N}^{\longleftarrow} \Gamma^{\infty}_{+,\{j,\mu\}}(q^{j+\half})
	\cdot \prod_{0\leq i<L}^{\longrightarrow} \Gamma^{\infty}_{-,\{i,\mu\}}(q^{i+\half})
	|\emptyset\rangle.
\end{align}
By virtue of the above formula,
we will prove this theorem by induction on the Frobenius length of $\mu$, i.e. $r$ if $\mu=(m_1,...,m_r|n_1,...,n_r)$ under the Frobenius notation.

At first, if the Frobenius length of $\mu$ is $0$, i.e. $\mu=\emptyset$,
this theorem holds since
\begin{align*}
	\tilde{Z}^{N+1,\infty,L+1}_{\emptyset,\emptyset,\emptyset}
	=\langle0|
	\prod_{0\leq j<N}\Gamma_+(q^{j+1/2})
	\cdot \prod_{0\leq i<L}\Gamma_-(q^{i+1/2})
	|0\rangle
	=\prod_{1\leq n\leq N \atop 1\leq l\leq L}
	(1-q^{n+l-1})^{-1}
\end{align*}
by the communication relations \eqref{eqn:gamma+- comm relation} of $\Gamma_{\pm}(\cdot)$.
This is a special case of the full MacMahon formula \eqref{eqn:macmahon}.

Now, we assume that the Frobenius length of $\mu$ is $r>0$,
and this theorem already holds for all $\tilde\mu$ whose Frobenius length is less than $r$.
In this case,
equation \eqref{eqn:z limit = vev} can be written as
\begin{align*}
	Z^{N+1,\infty,L+1}_{\emptyset,\mu,\emptyset}
	=\langle0|
	\prod_{m_1<i<N}\Gamma_+(q^{i+1/2})
	\cdot \Gamma_-(q^{-m_1-1/2})
	\cdots
	\Gamma_+(q^{-n_1-1/2})
	\cdot \prod_{n_1<i<L}\Gamma_-(q^{i+1/2})
	|0\rangle,
\end{align*}
where the omitted terms $\cdots$ are determined by $(m_2,...,m_r|n_2,...,n_r)$.
We replace the terms $\Gamma_-(q^{-m_1-1/2})$ and $\Gamma_+(q^{-n_1-1/2})$ appeared in the right hand side of above equation by
\begin{align}\label{eqn:replacement 1}
	\Gamma_-(q^{-m_1-1/2})\Gamma_+(q^{m_1+1/2})^{-1}
	\cdot\Gamma_+(q^{m_1+1/2})
\end{align}
and
\begin{align}\label{eqn:replacement 2}
	\Gamma_-(q^{n_1+1/2})
	\cdot\Gamma_-(q^{n_1+1/2})^{-1} \Gamma_+(q^{-n_1-1/2})
\end{align}
respectively.
Then, we can apply the commutation relations \eqref{eqn:gamma+- comm relation} to move terms in
\[\Gamma_-(q^{-m_1-1/2})\Gamma_+(q^{m_1+1/2})^{-1}
\quad\quad\text{and}\quad\quad
\Gamma_-(q^{n_1+1/2})^{-1}\Gamma_+(q^{-n_1-1/2})\]
to the leftmost or rightmost sides, depending on it is of the form $\Gamma_-(\cdot)$ or $\Gamma_+(\cdot)$.
We do this since $\Gamma_-(\cdot)$ and $\Gamma_+(\cdot)$ preserve the left and right vacuums respectively (see equation \eqref{eqn:gamma preserve vacuum})
and by doing that,
we can apply the induction process.

The result is that,
the commutation relations involved with $\Gamma_-(q^{-m_1-1/2})$ and $\Gamma_+(q^{-n_1-1/2})$ produce terms
\begin{align}\label{eqn:induction 12}
	\prod_{i=0}^{N-m_1-2} \frac{1}{1-q^{i+1}}
	\quad\quad\text{and}\quad\quad
	\prod_{i=0}^{L-n_1-2} \frac{1}{1-q^{i+1}}
\end{align}
respectively.
The commutation relations involved with $\Gamma_+(q^{m_1+1/2})^{-1}$ and $\Gamma_-(q^{n_1+1/2})^{-1}$ produce terms
\begin{align}\label{eqn:induction 3}
	\prod_{i=2}^r (1-q^{-m_i+m_1})
	\prod_{i=1}^r (1-q^{n_i+m_1+1})^{-1}
	\prod_{i=m_1}^{L+m_1-1} (1-q^{i+1})
\end{align}
and
\begin{align}\label{eqn:induction 4}
	\prod_{i=2}^r (1-q^{-n_i+n_1})
	\prod_{i=2}^r (1-q^{m_i+n_1+1})^{-1}
	\prod_{i=n_1}^{N+n_1-1} (1-q^{i+1})
\end{align}
respectively.
Denote by $\tilde{\mu}=(m_2,...,m_r|n_2,...,n_r)$.
We thus have,
the difference between $\tilde{Z}^{N+1,\infty,L+1}_{\emptyset,\mu,\emptyset}$
and $\tilde{Z}^{N+1,\infty,L+1}_{\emptyset,\tilde{\mu},\emptyset}$
is a factor consisting of multiplication of equations \eqref{eqn:induction 12}, \eqref{eqn:induction 3} and \eqref{eqn:induction 4}.
That is to say,
\begin{align}
	\tilde{Z}^{N+1,\infty,L+1}_{\emptyset,\mu,\emptyset}
	=&\tilde{Z}^{N+1,\infty,L+1}_{\emptyset,\tilde\mu,\emptyset}
	\cdot \prod_{i=N-m_1-1}^{N+n_1-1} (1-q^{i+1})
	\prod_{i=L-n_1-1}^{L+m_1-1} (1-q^{i+1})
	\label{eqn:c term}\\
	&\cdot \prod_{i=0}^{m_1-1} (1-q^{i+1})^{-1}
	\prod_{i=2}^r (1-q^{-m_i+m_1})
	\prod_{i=1}^r (1-q^{m_1+n_i+1})^{-1} \label{eqn:h m term}\\
	&\cdot \prod_{i=0}^{n_1-1} (1-q^{i+1})^{-1}
	\prod_{i=2}^r (1-q^{-n_i+n_1})
	\prod_{i=2}^r (1-q^{m_i+n_1+1})^{-1} \label{eqn:h n term}
\end{align}

By comparing the difference between the boxes in $\mu$ and $\tilde{\mu}$,
\begin{align*}
	\{(i,j)\in\mu\}
	=&
	\{(i+1,j+1)|(i,j)\in\mu'\}\\
	&\sqcup \{(1,j)|1\leq j\leq m_1+1\}
	\sqcup \{(i,1)|2\leq i\leq n_1+1\}.
\end{align*}
Recall that the content of a box in a Young diagram is $c(i,j)=j-i$,
so we have
\begin{align*}
	\{c(i,j)|(i,j)\in\mu\setminus\tilde\mu\}
	=\{m_1,m_1-1,...,0,...,-n_1\}.
\end{align*}
As a consequence,
the last two terms in the right hand side of equation \eqref{eqn:c term} can be rewritten as
\begin{align*}
	\prod_{(i,j)\in\mu\setminus\tilde\mu} (1-q^{N-c(i,j)})
	(1-q^{L+c(i,j)}).
\end{align*}
Similarly,
recall that the hook-length of a box in the Young diagram $\mu$ is defined by $h(i,j)=\mu_i+\mu^t_j-i-j+1$,
so when $1\leq i \leq r$,
we have
\begin{align*}
	h(1,i)=m_1+n_i-1
	\quad\quad\quad\text{and}\quad\quad\quad
	h(i,1)=m_i+n_1-1.
\end{align*}
On the other hand, we have
\begin{align*}
	\{h(1,i)|r<i\leq m_1+1\}
	=\{1,2,...,m_1\}\setminus\{m_1-m_2,m_1-m_3,...,m_1-m_r\}
\end{align*}
and
\begin{align*}
	\{h(i,1)|r<i\leq m_1+1\}
	=\{1,2,...,n_1\}\setminus\{n_1-n_2,n_1-n_3,...,n_1-n_r\}.
\end{align*}
Thus,
the equations \eqref{eqn:h m term} and \eqref{eqn:h n term} can be rewritten as
\begin{align*}
	\prod_{(i,j)\in\mu\setminus\tilde\mu} \frac{1}{1-q^{h(i,j)}}.
\end{align*}

In conclusion,
we have proven that
\begin{align*}
	\tilde{Z}^{N+1,\infty,L+1}_{\emptyset,\mu,\emptyset}
	=&\tilde{Z}^{N+1,\infty,L+1}_{\emptyset,\tilde\mu,\emptyset}
	\cdot \prod_{(i,j)\in\mu\setminus\tilde\mu} \frac{(1-q^{N-c(i,j)})
		(1-q^{L+c(i,j)})}{1-q^{h(i,j)}}.
\end{align*}
Thus,
the inductive hypothesis can be used and the equation \eqref{eqn:Z^(N,infty,L)} is proved.
$\Box$

\begin{rmk}
	It is worth mentioning again that,
	the right hand side of our formula \eqref{eqn:Z^(N,infty,L) main} is exactly equivalent to the expression of the open string amplitude of the double-$\mathbb{P}^1$ model with one nontrivial representation up to a MacMahon factor \cite{AKMV,LLLZ,IK06} (see the Corollary \ref{cor:dP1}).
	The general philosophy behind here is that,
	adding a wall to the crystal melting model is equivalent to gluing a new topological vertex (see \cite{Oku05,S06}).
	For the resolved conifold case, see \cite{Oku05} for a physical proof
	and for the closed topological case, see \cite{S06}.
	So, our formula \eqref{eqn:Z^(N,infty,L) main} should be explained as an open string amplitude version for the results in \cite{Oku05,S06}.
	This is our original motivation to prove this formula.
\end{rmk}

\begin{rmk}
	It was shown by Okounkov and Reshetikhin \cite{OR07} that the random skew plane partition model is a special Schur process.
	Then they obtained the corresponding partition function of this model (see the second equation in Theorem 2 in \cite{OR07}).
	As explained in the introduction of this paper,
	our $\tilde{Z}^{N+1,\infty,L+1}_{\emptyset,\mu,\emptyset}$ should be exactly equal to their partition function of the skew plane partitions,
	thus our formula \eqref{eqn:Z^(N,infty,L) main} should be equivalent to their formula even though they look different.
	
	Our formula \eqref{eqn:Z^(N,infty,L) main} mainly consists of three parts,
	the first part is a restriction condition for $\mu_1$ and $l(\mu)$,
	the second part is a special case of the full MacMahon formula \eqref{eqn:macmahon}
	and the last part is a product of terms labeled by boxes in $\mu$.
	This formula is efficient when comparing it to the open string amplitude of the double-$\mathbb{P}^1$ model and giving a new derivation of the full MacMahon formula.
	The formula by Okounkov and Reshetikhin (the second equation in Theorem 2 in \cite{OR07}) is a product of terms labeled by elements in two sets.
\end{rmk}

\begin{rmk}
	It will be an interesting problem to give a direct combinatorial proof of our formula \eqref{eqn:Z^(N,infty,L) main} about the partition function of plane partitions with a limit boundary.
	See, for examples, Theorem 7.20.1 in \cite{S99} for a bi-jective proof for the plane partitions without height restriction and Section 3 in \cite{V09} for the strict plane partitions.
\end{rmk}

\subsection{A new proof of the full MacMahon formula}
In this subsection,
by using the Theorem \ref{thm:main ex},
we give a new proof of the full MacMahon formula,
which gives the partition function of the plane partitions restricted in a box.

The very first proof of the full MacMahon formula can be seen in \cite{Mc} by MacMahon.
Later,
directly using Schur functions to obtain this formula was obtained by Stanley (see Theorem 7.21.7 in \cite{S99}).
By virtue of the methods of the Schur process and the crystal melting model,
in \cite{OR03} and \cite{ORV}, they gave a new proof of the MacMahon formula only for the $N=\infty$ case.
This is because dealing with the $z$-axis direction is much more complicated than the other two directions.
For example,
in Section 3.4 in \cite{ORV},
especially in their equation (3.16),
they said that they should let $N_3=\infty$ (which corresponds to $N=\infty$ in our paper) from the very beginning for obtaining the vacuum expectation value formula.
Also, the similar method was used in \cite{FW07,V07,V09} to study other types of plane partitions.
They generalized the MacMahon formula to the shifted plane partitions and the $(q,t)$-deformed case.
However,
all their results are stated for $N=\infty$,
even for the original MacMahon formula case,
which corresponds to partition functions of plane partitions without height restriction.
It is all because from the formula in terms of vacuum expectation value,
the three directions look very different
and the $z$-axis direction looks much more complicated than the other two directions,
even though they should be symmetric via the rotation and taking mirror of the plane partitions.
Our approach to derive the full MacMahon formula is motivated by the gluing rule of the topological vertex.
To be precise,
we use the symmetry of plane partitions with perpendicular type boundaries to transfer the height restriction to glue a new topological vertex,
and use our formula \eqref{eqn:Z^(N,infty,L) main} to deal with the terms that appeared in the gluing rule.

First,
we need the following two lemmas.
\begin{lem}\label{lem:<mu-1>}
	\begin{align}\label{eqn:<mu-1>}
		\langle\mu|
		\prod_{m=M}^{\infty} \Gamma_-(q^{m+1/2})^{-1}|0\rangle
		=q^{M|\mu|}\cdot s_{\mu^t}(x_m=-q^{m-1/2}).
	\end{align}
\end{lem}
{\bf Proof:}
First,
by taking dual,
the left hand side of equation \eqref{eqn:<mu-1>} is equal to
\begin{align*}
	\langle\mu|
	\prod_{m=M}^{\infty} \Gamma_-(q^{m+1/2})^{-1}|0\rangle
	=\langle0|
	\prod_{m=1}^{\infty} \Gamma_+(q^{m+M-1/2})^{-1}|\mu\rangle.
\end{align*}
Thus, by 
$\Gamma_+(z)^{-1}=\exp\Big(-\sum_{n=1}^\infty \frac{z^n \alpha_{n}}{n}\Big)$,
and equation \eqref{eqn:inf gamma+ action},
it is further equal to
\begin{align}\label{eqn:ex second part}
	\begin{aligned}
		(s_{\mu}|_{p_k\rightarrow-p_k})|_{x_m\rightarrow q^{m+M-1/2}}
		=&(s_{\mu}|_{p_k\rightarrow(-1)^{k-1}p_k})|_{x_m\rightarrow -q^{m+M-1/2}}\\
		=&s_{{\mu^t}}(x_m=-q^{m+M-1/2})\\
		=&q^{M|\mu|}\cdot s_{{\mu^t}}(x_m=-q^{m-1/2}),
	\end{aligned}
\end{align}
where the first and last equal signs come from $\deg s_{\mu}= \deg s_{\mu^t}=|\mu|$ and $\deg p_k=k$ if we assign $\deg x_m=1$ for all $1\leq m<\infty$.
In the second equal sign,
we used the involution $p_k\mapsto (-1)^{k-1} p_k$ for $1\leq k<\infty$
in the ring $\mathbb{C}[p_1,p_2,...]$
and the effect of this involution on Schur functions is equivalent to taking the transpose of partitions (see equation (3.8) in I.3 in \cite{Mac}). 
$\Box$

\begin{lem}\label{lem:cauchy}
	For any positive integers $M,N,L$,
	we have
	\begin{align}\label{eqn:g Cauchy in q}
		\sum_{\mu} s_{\mu}(q^{-\rho}|_L)
		\cdot q^{M|\mu|}s_{\mu^t} (-q^{-\rho}|_N)
		=\prod_{1\leq n\leq N \atop 1\leq l\leq L}
		(1-q^{n+l+M-1}),
	\end{align}
	where the sum is over all partitions and we recall that $q^{-\rho}|_N=(q^{1/2},q^{3/2},...,q^{(2N-1)/2},0,...)$.
\end{lem}
{\bf Proof:}
This lemma can be proved directly in terms of the Cauchy identity.
In terms of the following kind of Cauchy identity
(see equation (4.3') in I.4 in \cite{Mac})
\begin{align*}
	\sum_{\mu} s_{\mu}(x_1,x_2,...) s_{\mu^t}(y_1,y_2,...)
	=\prod_{i,j} (1+x_i y_j),
\end{align*}
the left hand side of equation \eqref{eqn:g Cauchy in q} is equal to
\begin{align*}
	\sum_{\mu} s_{\mu}(q^{(M-\rho)}|_L)
	\cdot s_{\mu^t} (-q^{-\rho}|_N)
	=&\prod_{1\leq n\leq N \atop 1\leq l \leq L}
	(1+x_ny_l)|_{x_n \rightarrow q^{M+l-1/2}
		\atop y_l \rightarrow -q^{n-1/2}},
\end{align*}
which is exactly equal to the right hand side of equation \eqref{eqn:g Cauchy in q}.
$\Box$

Recall the explanation in Remark \ref{rmk:empty = wall},
the diagonal empty condition is equivalent to adding a wall at the corresponding position.
Thus,
$\tilde{Z}^{L+1,N+1,M+1}_{\emptyset,\emptyset,\emptyset}$
is the partition function of all finite plane partitions in the region $[0,L]\times[0,N]\times[0,M]$
and the full MacMahon formula is then
\begin{align}\label{eqn:macmahon LNM}
	\tilde{Z}^{L+1,N+1,M+1}_{\emptyset,\emptyset,\emptyset}=
	\prod_{l=1}^L \prod_{n=1}^N \frac{1-q^{l+n+M-1}}{1-q^{l+n-1}}.
\end{align}

{\bf Proof of the full MacMahon formula:}
First,
by equation \eqref{eqn:main tZ} in Theorem \ref{thm:main},
\begin{align}
	\tilde{Z}^{L+1,N+1,M+1}_{\emptyset,\emptyset,\emptyset}
	=&\langle0|
	\prod_{n=0}^{L-1} \Gamma_+(q^{n+1/2})
	\mathbbm{1}_{l(\cdot^t)\leq N}
	\prod_{m=0}^{M-1} \Gamma_-(q^{m+1/2})
	0|\rangle\\
	=&\sum_{\mu}
	\langle0|
	\prod_{n=0}^{L-1} \Gamma_+(q^{n+1/2})
	\mathbbm{1}_{l(\cdot^t)\leq N}
	\prod_{m=0}^{\infty} \Gamma_-(q^{m+1/2})
	|\mu\rangle \label{eqn:pf 1part}\\
	&\quad\quad\quad\cdot\langle\mu|
	\prod_{m=M}^{\infty} \Gamma_-(q^{m+1/2})^{-1}|0\rangle
	\label{eqn:pf 2part}.
\end{align}
For equation \eqref{eqn:pf 1part},
it gives the partition function of diagonal plane partitions in the set $\tilde{\mathcal{P}}^{L+1,N+1,\infty}_{\emptyset,\emptyset,\mu}$,
so it is equal to $\tilde{Z}^{L+1,N+1,\infty}_{\emptyset,\emptyset,\mu}$ up to a global factor $q^{\binom{\mu^t}{2}+|\mu|/2}$ from Theorem \ref{thm:main}.
Meanwhile, $\tilde{Z}^{L+1,N+1,\infty}_{\emptyset,\emptyset,\mu}$ can be calculated in terms of Corollary \ref{cor:rotation} and Theorem \ref{thm:main ex}.
That is to say, equation \eqref{eqn:pf 1part} is equal to
\begin{align}\label{eqn:ex first part}
	\delta_{\mu_1\leq N}\prod_{1\leq n\leq N \atop 1\leq l\leq L}
	(1-q^{n+l-1})^{-1}
	\cdot \sum_{\mu} s_{\mu}(q^{-\rho}|_L)
	\cdot \prod_{(i,j)\in\mu^t} (1-q^{N+c(i,j)}),
\end{align}
where we have used the facts that $n(\mu)=\binom{\mu^t}{2}$ and
\begin{align*}
	s_{\mu}(q^{-\rho}|_L)
	=\delta_{l(\mu)\leq L}
	\cdot q^{n(\mu)+|\mu|/2}
	\cdot \prod_{(i,j)\in\mu}\frac{1-q^{L+c(i,j)}}{1-q^{h(i,j)}}.
\end{align*}
For equation \eqref{eqn:pf 2part},
it is equal to the equation in Lemma \ref{lem:<mu-1>}.
Thus, we obtain that $\tilde{Z}^{L+1,N+1,M+1}_{\emptyset,\emptyset,\emptyset}$ is equal to
\begin{align}\label{eqn:Z last step}
	\prod_{1\leq n\leq N \atop 1\leq l\leq L}
	(1-q^{n+l-1})^{-1}
	\cdot \sum_{\mu} s_{\mu}(q^{-\rho}|_L)
	\cdot\delta_{l(\mu^t)\leq N}
	q^{M|\mu|}s_{\mu^t}(-q^{-\rho})
	\cdot\prod_{(i,j)\in\mu^t} (1-q^{N+c(i,j)}).
\end{align}
By using the following equation
\begin{align*}
	\delta_{l(\mu^t)\leq N}
	\prod_{(i,j)\in\mu^t} (1-q^{N+c(i,j)})
	\cdot s_{\mu^t}(-q^{-\rho})
	=s_{\mu^t}(-q^{-\rho}|_N),
\end{align*}
which is obtained by comparing equations \eqref{eqn:schur q^rho} and \eqref{eqn:schur q^rho|N},
one can apply the Lemma \ref{lem:cauchy} to further simplify equation \eqref{eqn:Z last step} to obtain
\begin{align*}
	\tilde{Z}^{L+1,N+1,M+1}_{\emptyset,\emptyset,\emptyset}
	=\prod_{1\leq n\leq N \atop 1\leq l\leq L}
	(1-q^{n+l-1})^{-1}
	\cdot
	\prod_{1\leq n\leq N \atop 1\leq l\leq L}
	(1-q^{n+l+M-1}),
\end{align*}
which is equivalent to the full MacMahon formula \eqref{eqn:macmahon LNM}.
$\Box$

\begin{rmk}
	The above proof is similar to the method used in \cite{S06} to obtain a crystal melting model for the closed topological vertex.
	In Section 3.2 in \cite{S06}, he directly used the full MacMahon formula and showed that,
	up to a MacMahon factor,
	it equals the closed string amplitude of the closed topological vertex via the gluing rule.
	Our above proof reverses his method.
\end{rmk}

\section{Symmetric plane partitions with a limit boundary along $z$-axis direction}
\label{sec:case2}
In this section,
we obtain a product formula for the partition function of symmetric plane partitions bounded by two walls and possessing a limit boundary along $z$-axis direction.

First,
we review the definition of symmetric plane partitions (see \cite{Mac,S99}).
A plane partition $\pi$ is called symmetric if $\pi_{i,j}=\pi_{j,i}$ for all $i,j$.
Intuitively,
The 3D diagram of $\pi$ is mirror symmetric about the $(x-y=0)$ plane.
Similarly,
a partition $\mu$ is symmetric if $\mu=\mu^t$.
That is to say,
the Young diagram corresponding to $\mu$ is invariant under transpose.
For example,
if a symmetric plane partition $\pi$ has a limit boundary described by $\mu$ along the $z$-axis direction,
then $\mu$ must be symmetric.

We are mainly interested in the symmetric plane partitions bounded by two walls along $x$ and $y$-axis directions and possessing a limit boundary along the $z$-axis direction.
Alternatively, the skew symmetric partitions can also be considered equivalent.
To be precise,
we consider the following set of a special kind of symmetric plane partitions
\begin{align*}
	SP(N,\mu)=\{\pi\in\tilde{\mathcal{P}}^{N,\infty,N}_{\emptyset,\mu,\emptyset},
	\pi_{i,j}=\pi_{j,i}\}.
\end{align*}
Intuitively,
they are symmetric plane partitions bounded by two walls at $x=N, y=N$,
and has a limit boundary described by $\mu$ along the $z$-axis direction.

Unless $\mu=\emptyset$,
the plane partition in the set $SP(N,\mu)$ is not finite,
so we also need to use the modified size introduced in Subsection \ref{sec:diagonal boundary}.
That is to say,
we are interested in the following partition function
\begin{align}
	SZ(N,\mu)=
	\sum_{\pi\in SP(N,\mu)}
	q^{\widetilde{|\pi|}}.
\end{align}
Similar to the Subsection \ref{sec:diagonal boundary},
this partition function can also be regarded as a limit of the partition functions of symmetric plane partitions with height restriction.

The rest of this section is devoted to proving the product formula \eqref{eqn:SZ formula main} for the partition function $SZ(N,\mu)$.
First,
we need the following lemmas.
The first lemma is essentially the I.5.Example 4 in \cite{Mac}.
\begin{lem}
	We have
	\begin{align}\label{eqn:sum sla}
		\sum_{\lambda} s_{\lambda}=
		 \exp\Big(\sum_{n=1}
		 \frac{1}{n}(p_n+\frac{1}{2}p_n^2-\frac{1}{2}p_{2n})\Big),
	\end{align}
	where both sides of the above equation could be regarded as a formal power series in the ring $\mathbb{C}[\![p_1,p_2,...]\!]$
	or formal symmetric functions with respect to $\{x_i\}_{i=1}^\infty$.
\end{lem}
{\bf Proof:}
By the I.5.Example 4 in \cite{Mac},
we first obtain
\begin{align}\label{eqn:sum sla in mac}
	\sum_{\lambda} s_{\lambda}(x_1,x_2,...)=
	\prod_{i}(1-x_i)^{-1}
	\prod_{i<j}(1-x_ix_j)^{-1}.
\end{align}
Thus,
we only need to show that the right hand side of equation \eqref{eqn:sum sla in mac} is equal to the right hand side of equation \eqref{eqn:sum sla},
which can be finished just by using the definition of the $n$-power sum coordinates $p_n=p_n(x_1,x_2,...)$.
More precisely, notice that
\begin{align*}
	\log \prod_{i}(1-x_i)^{-1}
	=\sum_{n=1} \frac{1}{n}p_n
\end{align*}
and
\begin{align*}
	\log \prod_{i<j}(1-x_ix_j)^{-1}
	=\sum_{n=1}
	\frac{1}{2n}(p_n^2-p_{2n}).
\end{align*}
$\Box$

In terms of the notations in Subsection \ref{sec:pre schur},
the above lemma can also be rewritten as
\begin{align}\label{eqn:sum sla |la>}
	\sum_{\lambda} |\lambda\rangle=
	\exp\Big(\sum_{n=1}
	\frac{1}{n}(\alpha_{-n}+\frac{1}{2}\alpha_{-n}^2-\frac{1}{2}\alpha_{-2n})\Big)
	\cdot 1,
\end{align}
where the right hand side of the above equation should be regarded as the action of the operator,
multiplying the corresponding function,
on the constant function $1$.

\begin{lem}\label{lem: gamma+- sla}
	The operators $\Gamma_-(z)$ and $\exp\Big(\sum_{n=1}
	\frac{1}{n}(\alpha_{-n}+\frac{1}{2}\alpha_{-n}^2-\frac{1}{2}\alpha_{-2n})\Big)$
	commute with each other.
	For $\Gamma_+(z)$,
	we have the following commutation relation
	\begin{align}\label{eqn: gamma+ sla}
		\begin{aligned}
		\Gamma_+(z) \exp\Big(\sum_{n=1}&
		\frac{1}{n}(\alpha_{-n}+\frac{1}{2}\alpha_{-n}^2-\frac{1}{2}\alpha_{-2n})\Big) \\
		=&\frac{1}{1-z}\exp\Big(\sum_{n=1}
		\frac{1}{n}(\alpha_{-n}+\frac{1}{2}\alpha_{-n}^2-\frac{1}{2}\alpha_{-2n})\Big)
		\Gamma_-(z)\Gamma_+(z).
		\end{aligned}
	\end{align}
\end{lem}
{\bf Proof:}
The first result that $\Gamma_-(z)$ and $\exp\Big(\sum_{n=1}
\frac{1}{n}(\alpha_{-n}+\frac{1}{2}\alpha_{-n}^2-\frac{1}{2}\alpha_{-2n})\Big)$
commute with each other trivially follows from the fact that both of them only consist of $\alpha_{-n}$ for $n>0$.
The equation \eqref{eqn: gamma+ sla} follows from the following three commutation relations
\begin{align*}
	\Gamma_+(z)
	\exp\Big(\sum_{n=1}
	\frac{1}{n}\alpha_{-n}\Big)
	=\frac{1}{1-z}\exp\Big(\sum_{n=1}
	\frac{1}{n}\alpha_{-n}\Big)
	\Gamma_+(z),
\end{align*}
\begin{align*}
	\Gamma_+(z)\exp\Big(\sum_{n=1}
	-\frac{1}{2n}\alpha_{-2n}\Big)
	=(1-z^2)^{1/2}\exp\Big(\sum_{n=1}
	-\frac{1}{2n}\alpha_{-2n}\Big)
	\Gamma_+(z),
\end{align*}
\begin{align*}
	\Gamma_+(z)\exp\Big(\sum_{n=1}
	\frac{1}{2n}\alpha_{-n}^2\Big)
	=(1-z^2)^{-1/2}\exp\Big(\sum_{n=1}
	\frac{1}{2n}\alpha_{-n}^2\Big)
	\Gamma_-(z)\Gamma_+(z).
\end{align*}
All of the above three equations can be proved by using Baker--Campbell--Hausdorff formula
and for the last equation,
we also need the following
\begin{align*}
	\exp\Big(\sum_{n=1}
	\frac{z^n}{n}(\alpha_n+\alpha_{-n})\Big)
	=(1-z^2)^{-1/2}
	\Gamma_-(z)\Gamma_+(z),
\end{align*}
which is proved by the Zassenhaus formula.
$\Box$

\begin{thm}[= Theorem \ref{thm:SZ main}]
	The partition function $SZ(N+1,\mu)$ of symmetric plane partitions with a limit boundary has the following product formula
	\begin{align}\label{eqn:SZ formula}
		\begin{aligned}
		SZ(N+1,\mu)=&
		\delta_{\mu_1\leq N}
		\cdot\prod_{i=0}^{N-1}\frac{1}{(1-q^{2i+1})\prod_{j=0}^{i-1}(1-q^{2(i+j+1)})}\\
		&\quad\quad\quad\quad\cdot \frac{\prod_{(i,j)\in\mu}(1-q^{2N+2c(i,j)})}
		{\prod_{(i,i)\in\mu}(1-q^{h(i,i)})
		\prod_{(i,j)\in\mu\atop i<j}(1-q^{2h(i,j)})},
	\end{aligned}
	\end{align}
	where $\mu$ is a symmetric partition,
	$\delta_{\mu_1\leq N}=1$ if $\mu_1\leq N$ and otherwise it is $0$.
\end{thm}
{\bf Proof:}
If $\mu_1>N$,
the set $SP(N+1,\mu)$ is empty,
and thus $SZ(N+1,\mu)=0$.

From now on, we assume $\mu_1\leq N$.
First, 
similar to the proof in Theorem \ref{thm:main},
by deleting the cubes in the region $\{(x,y,z)|y\leq \mu^t_{\lfloor x\rfloor+1}\text{\ for\ } 0\leq x<N+1\}$,
one thus transfer $\pi\in SP(N+1,\mu)$ to be a skew plane partition $\pi'$ in the sense of \cite{OR07}.
By dividing $\pi'$ along the diagonal slices,
there is a one-to-one correspondence between it and a series of interlacing Young diagrams $(\mu^k)_{k=-N-1}^{N+1}$ satisfying the following conditions:

i) $\mu^{-N-1}=\mu^{N+1}=\emptyset$,
$\mu^{-i}=\mu^i$ for all $i$,

ii) $\mu^{j-1}\ (\prec)_{j,\mu}\ \mu^{j}$ for $-N<j\leq0$,
and $\mu^{i}\ (\succ)_{i,\mu}\ \mu^{i+1}$ for $0\leq i<N$.\\
Since $\mu^{-i}=\mu^i$,
we can only record the $\{\mu^k\}_{k=-N-1}^0$ parts.
Then,
the method used in Section \ref{sec:case1} (see \cite{ORV}) gives
\begin{align*}
	SZ(N+1,\mu)
	=&\sum_{\{\mu^k\}_{k=-N-1}^{0}
	\text{\ satisfying\ certain\ conditions}}
	q^{2\sum_{k=-N}^{-1}|\mu^k|+|\mu^0|}\\
	=&\sum_{\lambda}
	\langle0|\prod_{0\leq j<N}^{\longleftarrow} \big(q^{2L_0}\cdot \Gamma^{\infty}_{+,\{j,\mu\}}(1)\big)
	\cdot q^{L_0}|\lambda\rangle\\
	=&\sum_{\lambda}
	\langle0|\prod_{0\leq j<N}^{\longleftarrow} \Gamma^{\infty}_{+,\{j,\mu\}}(q^{2j+1})
	|\lambda\rangle.
\end{align*}
From equation \eqref{eqn:sum sla |la>},
we also have
\begin{align*}
	SZ(N+1,\mu)
	=\langle0|
	\prod_{0\leq j<N}^{\longleftarrow} \Gamma^{\infty}_{+,\{j,\mu\}}(q^{2j+1})
	|\exp\Big(\sum_{n=1}
	\frac{1}{n}(\alpha_{-n}+\frac{1}{2}\alpha_{-n}^2-\frac{1}{2}\alpha_{-2n})\Big)
	|0\rangle.
\end{align*}
As a consequence,
this theorem is equivalent to showing the equivalence of the right hand side of the above equation and the right hand side of equation \eqref{eqn:SZ formula}.
We will prove this fact below by induction on the Frobenius length $r(\mu)$ of $\mu$.

First,
For the $r(\mu)=0$ case, i.e. $\mu=\emptyset$,
we directly prove the following formula
\begin{align}\label{eqn:symm mu=0}
	\begin{aligned}
	\langle0|
	\prod_{i=0}^{N-1}\Gamma_+(q^{2i+1})
	|\exp\Big(\sum_{n=1}&
	\frac{1}{n}(\alpha_{-n}+\frac{1}{2}\alpha_{-n}^2-\frac{1}{2}\alpha_{-2n})\Big)
	|0\rangle\\
	&=\prod_{i=0}^{N-1}\frac{1}{(1-q^{2i+1})\prod_{j=0}^{i-1}(1-q^{2(i+j+1)})}.
	\end{aligned}
\end{align}
For the left hand side of the above equation,
we can move the term $\Gamma_+(q^{2N-1})$ to the rightmost side by using commutation relation \eqref{eqn: gamma+ sla}.
By the fact that $\Gamma_+(\cdot)$ preserves the vacuum vector $|0\rangle$,
the left hand side of equation \eqref{eqn:symm mu=0} is equal to
\begin{align*}
	\frac{1}{1-q^{2N-1}}\langle0|
	\prod_{i=0}^{N-2}\Gamma_+(q^{2i+1})
	|\exp\Big(\sum_{n=1}
	\frac{1}{n}(\alpha_{-n}+\frac{1}{2}\alpha_{-n}^2-\frac{1}{2}\alpha_{-2n})\Big)
	\Gamma_-(q^{2N-1})|0\rangle.
\end{align*}
Then,
by Lemma \ref{lem: gamma+- sla} and commutation relation \eqref{eqn:gamma+- comm relation},
we can move the term $\Gamma_-(q^{2N-1})$ in the above equation to the leftmost side and use the fact that $\Gamma_-(\cdot)$ preserves the dual vacuum vector $\langle0|$ to obtain,
the left hand side of equation \eqref{eqn:symm mu=0} is equal to
\begin{align*}
	\frac{1}{(1-q^{2N-1})
		\prod_{j=0}^{N-2}(1-q^{2N+2j})}\langle0|
	\prod_{i=0}^{N-2}\Gamma_+(q^{2i+1})
	|\exp\Big(\sum_{n=1}
	\frac{1}{n}(\alpha_{-n}+\frac{1}{2}\alpha_{-n}^2-\frac{1}{2}\alpha_{-2n})\Big)|0\rangle.
\end{align*}
One can notice that there are only $N-1$ $\Gamma_+(\cdot)$ in the above vacuum expectation value.
Thus, by repeating the above process,
equation \eqref{eqn:symm mu=0} is proved.
In conclusion,
we finish the proof of the $\mu=\emptyset$ case of this theorem.

Next,
we assume the Frobenius length of $\mu$ is $r>0$
and assume that this theorem already holds for any $\tilde{\mu}$ satisfying $r(\tilde\mu)<r$.
Since $\mu$ is symmetric,
we write $\mu=(m_1,...,m_r|m_1,...,m_r)$.
Then $SZ(N+1,\mu)$ is equal to
\begin{align*}
	\langle0|
	\prod_{m_1<i<N}\Gamma_+(q^{2i+1})
	\cdot \Gamma_-(q^{-2m_1-1})
	\cdots
	|\exp\Big(\sum_{n=1}
	\frac{1}{n}(\alpha_{-n}+\frac{1}{2}\alpha_{-n}^2-\frac{1}{2}\alpha_{-2n})\Big)|0\rangle,
\end{align*}
where the omitted terms $\cdots$ are determined by $(m_2,...,m_r)$.
For applying the induction process,
we replace the term $\Gamma_-(q^{-2m_1-1})$ appeared in the right hand side of above equation by
\begin{align}\label{eqn:replacement 1 sz}
	\Gamma_-(q^{-2m_1-1})\Gamma_+(q^{2m_1+1})^{-1}
	\cdot\Gamma_+(q^{2m_1+1}).
\end{align}
Then, we can apply the commutation relations \eqref{eqn:gamma+- comm relation} and \eqref{eqn: gamma+ sla} to move terms
$\Gamma_-(q^{-2m_1-1/2})$ and $\Gamma_+(q^{2m_1+1})^{-1}$
to the leftmost and rightmost sides, respectively.
The result is that,
\begin{align}
	S&Z(N+1,\mu)
	=\prod_{m_1<i<N}\frac{1}{(1-q^{2i-2m_1})}
	\cdot \prod_{i=2}^r (1-q^{2m_1-2m_i})
	\cdot \frac{1}{1-q^{2m_1+1}} \label{eqn:SZ gamma->1}\\
	&\cdot\langle0|
	\prod_{m_1<i<N}\Gamma_+(q^{2i+1})
	\Gamma_+(q^{2m_1+1})
	\cdots
	\exp\Big(\sum_{n=1}
	\frac{1}{n}(\alpha_{-n}+\frac{1}{2}\alpha_{-n}^2-\frac{1}{2}\alpha_{-2n})\Big)
	\Gamma_-(q^{2m_1+1})^{-1}|0\rangle \label{eqn:SZ gamma->}.
\end{align}
Once again,
we can apply the Lemma \ref{lem: gamma+- sla} and commutation relation \eqref{eqn:gamma+- comm relation} to move $\Gamma_-(q^{2m_1+1})^{-1}$ in the above equation to the leftmost side.
As a consequence,
the equation \eqref{eqn:SZ gamma->} is equal to
\begin{align*}
	\prod_{0\leq i<N}(1-q^{2i+2m_1+2})
	\cdot \prod_{i=2}^{r} \frac{1}{1-q^{2m_1+2m_i+2}}
	\cdot SZ(N+1,\tilde\mu),
\end{align*}
where $\tilde\mu=(m_2,...,m_r|m_2,...,m_2)$
whose Frobenius length is $r-1<r$.
Thus,
equations \eqref{eqn:SZ gamma->1} and \eqref{eqn:SZ gamma->} reduce to
\begin{align}
	SZ(N+1,\mu)
	=&SZ(N+1,\tilde{\mu})
	\cdot \prod_{i=-m_1}^{m_1}(1-q^{2N+2i}) \label{eqn:SZ c}\\
	&\cdot \prod_{i=0}^{m_1-1}\frac{1}{1-q^{2i+2}}
	\cdot \prod_{i=2}^r (1-q^{2m_1-2m_i})
	\cdot \frac{1}{1-q^{2m_1+1}}
	\cdot \prod_{i=2}^{r} \frac{1}{1-q^{2m_1+2m_i+2}} \label{eqn:SZ h}.
\end{align}
First,
the correspondence
\begin{align*}
	\{c(i,j)|(i,j)\in\mu\setminus\tilde\mu\}
	=\{-m_1,-m_1+1,...,0,...,m_1\}.
\end{align*}
tells that the second part in the right hand side of equation \eqref{eqn:SZ c} can be rewritten as
\begin{align}\label{eqn:SZ result c}
	\prod_{i=-m_1}^{m_1}(1-q^{2N+2i})
	=\prod_{(i,j)\in\mu\setminus\tilde\mu}
	(1-q^{2M+2c(i,j)}).
\end{align}
And on the other hand,
since $h(1,1)=2m_1+1,
h(1,i)=m_1+m_i+1$ for $2\leq i\leq r$
and
\begin{align*}
	\{h(1,i)|r<i\leq m_1+1\}
	=\{1,2,...,m_1\}\setminus\{m_1-m_2,m_1-m_3,...,m_1-m_r\},
\end{align*}
the equation \eqref{eqn:SZ h} is equal to
\begin{align}\label{eqn:SZ result h}
	\frac{1}{(1-q^{h(1,1)})
		\cdot\prod_{(1,j)\in\mu\atop 1<j} (1-q^{2h(1,j)})}
	=
	\frac{1}{\prod_{(i,i)\in\mu\setminus\tilde\mu}(1-q^{h(i,i)})
		\cdot\prod_{(i,j)\in\mu\setminus\tilde\mu\atop i<j} (1-q^{2h(i,j)})}.
\end{align}
By inserting equations \eqref{eqn:SZ result c} and \eqref{eqn:SZ result h}
into equations \eqref{eqn:SZ c} and \eqref{eqn:SZ h},
this theorem is thus proved by induction.
$\Box$

\begin{rmk}
	Borodin and Rains studied the free boundary Schur process \cite{BR05} (see also \cite{BBNV}),
	which is a Pfaffian analog of the original Schur process considered by Okounkov and Reshetikhin \cite{OR03}.
	They showed that it is a Pfaffian point process and obtained a formula for the partition function.
	Our method in this section is similar to theirs, and the usage of the equation \eqref{eqn:sum sla |la>} corresponds to their free boundary condition.
\end{rmk}

\vspace{.2in}
{\em Acknowledgements}.
The author is grateful to the anonymous referee for valuable suggestions.
The author would like to thank Professors Shuai Guo, Xiaobo Liu and Zhiyuan Wang for helpful discussions
and thank Professors Jian Zhou and Xiangyu Zhou for encouragement.
The author is supported by the NSFC grants (No. 12288201, 12401079),
the China Postdoctoral Science Foundation (No. 2023M743717),
and the China National Postdoctoral Program for Innovative Talents (No. BX20240407).

\vspace{.2in}

%%%%%%%%%%%%%%%%%%%%%%%%%%%%%%%%%%%%%%%%%%%%%%%
\renewcommand{\refname}{Reference}
\bibliographystyle{plain}
\bibliography{reference}
\vspace{30pt} \noindent
%%%%%%%%%%%%%%%%%%%%%%%%%%%%%%%%%%%%%%%%%%%%%%%%%%%%%%%%%%%%%%%%%%%%%%%%%%%%%%%%%%%
\end{document}